\documentclass[12pt,a4paper,oneside]{article}
\pdfoutput=1
\usepackage{graphicx,sectsty,amssymb,amsmath}
\usepackage[linkcolor={blue},citecolor={red},colorlinks=true,linktocpage]
{hyperref}
\usepackage[margin=2cm]{geometry}
\usepackage{multirow}

\usepackage{appendix}
\usepackage{amsmath}
\usepackage{tensor}
\usepackage{etoolbox}
\AtBeginEnvironment{align}{\setcounter{subeqn}{0}}
\newcounter{subeqn} %

\usepackage{setspace}
\usepackage{pdfpages}
\numberwithin{equation}{section}
\usepackage{cancel}
\usepackage{slashed}
\usepackage{xparse} 
\usepackage{mathtools} 
 \usepackage[nottoc]{tocbibind} 
\usepackage{titling} 
\usepackage[affil-it]{authblk} 

\newcommand{\beq}{\begin{equation}}
\newcommand{\eeq}{\end{equation}}

\newcommand{\pder}[2]{\frac{\partial#1}{\partial#2}} 

\DeclareMathOperator{\rk}{rk}

\DeclarePairedDelimiter\ceil{\lceil}{\rceil} 
\DeclarePairedDelimiter\floor{\lfloor}{\rfloor}

\title{Weak Coupling Limits and Colliding Punctures in Class-$\mathcal{S} $ Theories}
\author{Arel Genish
   \thanks{E-mail: \texttt{Arel.genish@weizmann.ac.il}} }
\author{Vladimir Narovlansky
   \thanks{E-mail: \texttt{Vladimir.narovlansky@weizmann.ac.il}} }
\affil{Department of Particle Physics and Astrophysics, \\ Weizmann Institute of Science, Rehovot 7610001, Israel}
\date{}

\makeatletter 
\let\@fnsymbol\@arabic
\makeatother

\begin{document}

\begin{titlingpage}
    \maketitle
    \begin{abstract}
Class-$\mathcal{S} $ theories are four-dimensional $\mathcal{N} =2$ supersymmetric field theories constructed by the reduction of a $(2,0)$ six-dimensional theory on a punctured Riemann surface $C$ (called the UV curve). A basic degeneration limit of the surface $C$ is when several punctures are brought close to each other. As this happens, a long tube appears in $C$ and a weakly coupled gauge group emerges. When the corresponding gauge coupling is turned off, we are left with two surfaces. This decoupling leads to a unique result. We explain how to fix the gauge group that becomes weakly coupled and the resulting two theories, in terms of the types of punctures on the surface $C$. Each of the resulting theories is given in terms of a class-$\mathcal{S} $ theory (plus possibly additional free hypermultiplets). Many other more complicated degeneration limits can be described by repeatedly using this specification.

 For questions about weak coupling limits in such theories, it is useful to consider the set of punctures that can appear at the end of a tube. These are the punctures that are created in the resulting surfaces when the gauge coupling is taken to be extremely weak. Not any puncture can be formed in a decoupling process, and this set is described. Each such puncture can be found diagrammatically in terms of the other punctures on $C$ in a particular decoupling limit. The relation of the gauge group to the symmetry associated with this puncture is discussed.
 
We discuss the relation of such degeneration limits of large $N$ theories, through the AdS/CFT correspondence, to decoupled field theories on $AdS_5$.

    \end{abstract}
\end{titlingpage}

\tableofcontents


\section{Introduction}

Consider the Coulomb branch of $\mathcal{N} =2$ superconformal field theories in four dimensions. At low energies we have $U(1)^n$ gauge fields, which are part of $\mathcal{N} =2$ vector multiplets. Denote the $\mathcal{N} =1$ chiral superfield part of these vector multiplets by $a_i$. In the low energy effective action, the gauge coupling constants matrix is promoted to $\tau^{ij} (a)$ (a function of the $a_i$). The gauge coupling matrix is related to the $\mathcal{N} =2$ prepotential $F$ by $\tau ^{ij} = \frac{\partial ^2 F}{\partial a_i \partial a_j} $, and we have the definition $a_D^i= \pder{F}{a_i} $. $a_D$ plays the role of $a$ in the electric-magnetic dual theory. Additionally, as can be observed from the low energy effective theory, the expectation values of $a_i$ and $a_D^i$ are the coefficients of the electric and magnetic charges under the $i$th $U(1)$ in the BPS bound.

The solution of the theory on the Coulomb branch (in the infrared) is described by the SW curve \cite{Seiberg:1994rs}\cite{Seiberg:1994aj}.
A basis of non-trivial cycles and dual cycles is chosen, and the integrals of the SW differential $\lambda $ along them give $a_i$ and $a_D^i$. The SW curve contains gauge invariant Coulomb branch parameters, and $a_i$, $a_D^i$ are holomorphic functions of these parameters.

The SW curve can actually be realized geometrically \cite{Witten:1997sc}. Linear quivers of $SU(k_i)$ gauge groups with $\mathcal{N} =2$ supersymmetry were constructed through the decoupling limit of configurations of NS5- and D4-branes in type IIA superstring. These configurations have singularities, but when interpreted in M-theory, could be described by a single M5-brane. For generic parameters, the M5-brane is smooth and contains no singularities. The description of the M5-brane includes the SW curve solution of the theory. Linear quivers with generic flavor groups of fundamental hypermultiplets of the $SU(k_i)$ gauge groups, are constructed by including D6-branes as well. Circular quivers, as well as other classes of $\mathcal{N} =2$ theories can also be obtained by brane constructions.

Following \cite{Gaiotto:2009we}, consider a linear quiver of $n$ $SU(N)$ gauge groups, with $N$ fundamentals at each end of the quiver. For $N=2$, when all the gauge groups but one are arbitrarily weakly coupled, we can take the coupling constant of that one gauge group to be very strong and apply the familiar $SL(2,\mathbb{Z} )$ S-duality of $SU(2)$ with four flavors. This leads to a generalized quiver (when expressed in terms of the weakly coupled gauge groups), differing from the linear quiver we began with. Similarly, for $N=3$, we may apply the Argyres-Seiberg duality (mentioned below), and get another kind of generalized quiver. These generalized quivers are composed of elementary building blocks. The $E_6$ SCFT is one of the building blocks for $N=3$. The generalized quivers constructed starting from some fixed quiver gauge theory, are weak coupling cusps of a single theory.

The linear quiver we started with can be constructed, as mentioned above, by taking $N$ M5-branes which are intersected by other $n+1$ transverse M5-branes. This configuration can be viewed as $N$ M5-branes wrapping a Riemann sphere, in the presence of $n+1$ punctures at some points on the Riemann sphere, as well as punctures at $0$ and $\infty $. In the circular quivers construction, the Riemann sphere is replaced by a torus $T^2$ (as a result of the topology of the space that the $N$ M5-branes wrap).

Other kinds of linear quivers will be described by including 2 punctures of a more general type. These punctures are local and thus suggest that we can have additional theories by combining different sets of the punctures. This more general sort of theories \cite{Gaiotto:2009we} is therefore defined by taking the $A_{N-1} $ $(2,0)$ six dimensional theory on a Riemann surface $C$, with co-dimension two defect operators (a twisting is required to get $\mathcal{N} =2$ in the four dimensional macroscopic theory). There are $(2,0)$ six dimensional theories of A,D and E types. This defines a four dimensional theory by a simply laced Lie group and a punctured Riemann surface. These theories are known as class-$\mathcal{S} $. We will restrict ourselves to the $A_{N-1} $ theories \footnote{For a review, see \cite{Tachikawa:2013kta}. Generalizations of our analysis to the D,E type theories is left for the future, see \cite{Chacaltana:2011ze},\cite{Chacaltana:2014jba}. }. 

A basic building block of the generalized quivers is the $T_N$ theory which can be identified by an $A_{N-1} $ theory on a sphere with 3 full punctures (a review of the types of punctures will be given below). It plays a similar role to that played by the $E_6$ SCFT in $N=3$ (even though there is another kind of isolated SCFTs which is identified more naturally with the generalization of the $E_6$ SCFT to general $N$). The $T_N$ theory is an example of a class-$\mathcal{S} $ theory with no brane construction of the sort described above. There is a description \cite{Benini:2009gi} of it in terms of a web of 5-branes (and 7 branes) in type IIB giving a five dimensional theory, which when compactified on $S^1$ gives the $T_N$ theory (as well as more general isolated SCFTs).  

The SW curve for the $A_{N-1} $ class-$\mathcal{S} $ theories can be written as an $N$ sheeted branched covering of $C$. It has the canonical form
\begin{equation} \label{eq:general_curve_form_class_S}
x^N+\sum _{k=2} ^N \phi _k(z) x^{N-k} =0 ,
\end{equation}
where $z$ parametrizes $C$ and the SW differential being $\lambda =x dz$. The $\phi _k$ are more naturally used as $k$-differentials $\phi _k dz^k$, having appropriate poles at the punctures. Usually the SW curve describes the infrared limit of the four dimensional theory on the Coulomb branch. Here, as is manifest in the brane constructions mentioned above, the structure of the SW curve as a branched covering of $C$ identifies the four dimensional theory. 

There are discrete holomorphic transformations keeping the SW curve with the covering structure above invariant, and are therefore symmetries of the full four dimensional theory. Then, the S-duality invariant space of exactly marginal deformations is the complex structure moduli space of $C$, with punctures of the same kind being indistinguishable.

At various cusps of the moduli space of the punctured Riemann surface $C$, the surface degenerates, long tubes emerge and some cycles shrink. In such cases, weakly coupled gauge groups emerge. A common situation where this happens, is when some punctures on $C$ are brought close to each other. At different degenerations of the same surface $C$, different gauge groups become weekly coupled. This provides us with S-dualities between a priori different theories. An example is the Argyres-Seiberg duality \cite{Argyres:2007cn}, stating that the strongly coupled $SU(3)$ superconformal theory with $6$ fundamental hypermultiplets is a weakly coupled $SU(2)$ theory, coupled to one fundamental hypermultiplet and the $T_3$ theory (which has global symmetry $E_6$ and no marginal deformations).

Let us write the form of the SW curve more explicitly for a few low genus surfaces $C$.
On a sphere (genus $g$=0) with punctures at $z_1,z_2, \dots $, $\phi _k$ of a massless theory is of the following form
\begin{equation}
\phi _k = \frac{Q_k(z)}{(z-z_1)^{p^1_k} (z-z_2)^{p_k^2} \dots } dz^k ,
\end{equation}
where we label by $p^i_k$ the pole structure corresponding to the puncture located at $z_i$.
Since we do not have a pole at infinity, a change of variable $z=1/w$ shows that the polynomial $Q_k$ is of order at most $\sum  _i p_k^i - 2k$. Therefore the number of Coulomb branch parameters $\phi _k$ gives rise to, is $\sum  _i p_k^i - 2k + 1$. (Note we could choose to position one of the poles at infinity, with the same result.) \\
On a torus ($g=1$), the general $\phi _k$ with the required pole structure (and no masses) is of the form
\begin{equation}
\phi _k = A_k \frac{\theta (z-n^1_k) \dots \theta (z-n^{d_k}_k)}{\theta (z-z_1)^{p_k^1} \theta (z-z_2)^{p_k^2} \dots } dz^k, \qquad d_k=\sum _i p_k^i, \qquad \sum _i n^i_k = \sum  _i p_k^i z_i ,
\end{equation}
where the $\theta $ is a Jacobi theta function. 
Including $A_k$ and the restriction on $\sum_i n^i_k$, we see that $\phi_k$ gives rise to $\sum _i p^i_k$ Coulomb branch parameters. In a general surface of genus $g$, the number of Coulomb branch parameters of dimension $k$ is the dimension of the space of $k$-differentials and is given by
\begin{equation} \label{eq:Coulomb_branch_graded_dimension}
d_k = \sum _i p^i_k + (g-1)(2k-1) .
\end{equation}

The pole structure of regular punctures in a superconformal class-$\mathcal{S} $ $A_{N-1} $ theory is restricted. A regular puncture $P$ is described in terms of a Young diagram with $N$ boxes in total. The pole structure of the puncture is fixed by the diagram as follows. For $k=2,\dots ,N$, $\phi _k$ has a pole of order $p_k$ at the puncture, where $p_k$ is given by $p_k=k-h(k,P)$, in which $h(k,P)$ is the row number of the $k$th box in the Young diagram (we label the rows starting with $1$). The most common punctures are simple and full punctures. A simple puncture has a diagram with rows of width $2,1,1,\dots $ and pole structure $1,1,1,\dots $ ($p_k=1$). A full puncture has a single row (of width $N$) and pole structure $1,2,3,\dots $ ($p_k=k-1$). Any Young diagram corresponds to a regular puncture, except for a single column diagram, referred to as a no-puncture.

Since we will commonly use punctures, let us introduce a convenient (but slightly subtle) notation.
Punctures will be denoted by upper case Roman letters, such as $P$. For each such puncture $P$ we will use the notations:
\begin{itemize}
\item $P_i$: the width of row number $i$.
\item $p_k$: the pole structure at the value $k$.
\item $p$: the number of boxes outside the first column (explicitly $p=\sum _i (P_i-1)$).
\item $h(k,P)$: the row number of box number $k$.
\end{itemize}
When we have several punctures, we add a superscript, as in $P^i$ (and then we have as before $P^i_j$, $p^i_k$, $p^i$ and $h(k,P^i)$).


Each regular puncture has a global symmetry associated with it. Corresponding to this symmetry, mass deformations can be introduced. The symmetry associated with a regular puncture $P$ will be denoted by $G(P)$ and is given by 
\begin{equation} \label{eq:regular_puncture_symmetry}
G(P) = S\left( \prod _i U(P_i - P_{i+1} ) \right)
\end{equation}
(where $S(\dots)$ means removing the diagonal $U(1)$).
The product of the punctures' symmetries does not have to be the full symmetry of the theory. A method to find the full symmetry which can be used in some of the cases is by considering the mirror of the theory compactified to three-dimensions, see \cite{Chacaltana:2010ks} using \cite{Benini:2010uu},\cite{Gaiotto:2008ak}.

In \autoref{sec:decoupling} we describe the result of decoupling in a general surface. We imagine that several punctures are brought close to each other, resulting in a formation of a long tube and a weakly coupled gauge group (see for instance \autoref{fig:sphere_decoupling}). In the extreme weak coupling limit, the Riemann surface is separated into two surfaces. We explain how, in a unique and simple manner, the result of this decoupling can be obtained. That is, what is the weakly coupled gauge group that arises, and what are the two theories corresponding to the two surfaces.

Then, in \autoref{sec:diagrammatic_decoupling} we first identify the punctures that are created at the ends of a tube as a set that will be useful to describe. The set of regular punctures that can appear in a decoupling process at the end of a tube is precisely the set of punctures $L$ satisfying $L_1 \ge 2L_2$, excluding the simple punctures of $N>2$.
 A simple diagrammatic method for finding these punctures in a given decoupling process is described.
There are theories that are formed after a decoupling which can be described in terms of irregular punctures, as reviewed in \autoref{sec:decoupling} following \cite{Chacaltana:2010ks},\cite{Chacaltana:2011thesis}. The set of irregular punctures in $\mathcal{N} =2$ SCFTs of the sort described there is classified in terms of Young diagrams.

In \autoref{sec:gauging} similar questions are asked from a different point of view, in which we consider what gauging of some regular puncture is possible. In other words, gauging a diagonal subgroup of the symmetry associated with a regular puncture and of a global symmetry from an additional theory, what are the possibilities for that additional theory that will result in a SCFT. The possible such gauge groups are listed. Additionally, we discuss the embedding of the gauged group in the symmetry associated with the puncture and a few implications of that.

The $\mathcal{N} =2$ circular quiver theory is studied using holography in \cite{Aharony:2015zea}. It turns out that in a certain limit and when $N$ is large, it contains the $\mathcal{N} =(2,0)$ six dimensional $A_{K-1}$ theory on $AdS_5 \times S^1$. In order to place the $(2,0)$ theory on $AdS_5 \times S^1$, boundary conditions must be specified. Two sorts of boundary conditions in this context are discussed in \cite{Aharony:2015zea} and will be mentioned here. It is natural to check whether there are other class-$\mathcal{S} $ theories in which there is a decoupled field theory on $AdS_5$. Additionally, for the $\mathcal{N} =2$ theories in which this is the case, various additional boundary conditions might be possible, implemented in other $\mathcal{N} =2$ theories analogous to the ones described in \cite{Aharony:2015zea}. These issues are discussed in \autoref{sec:large_N} and the conclusions of the other sections can be applied in this context.


\section{Weak coupling limits of a class-$\mathcal{S} $ theory} \label{sec:decoupling}

As was mentioned in the previous section, a common situation in which the Riemann surface $C$ of an $A_{N-1} $ class-$\mathcal{S} $ theory degenerates, is when several punctures are brought close to each other. When $C$ is a sphere, this is the only possibility for a degeneration. A long tube is then formed, and is associated with an emergent weakly coupled gauge group which we denote by $G_T$ (see \autoref{fig:sphere_decoupling}) . In the extreme weak coupling limit we are left with two surfaces describing two theories. We describe what are these theories for the different possible surfaces $C$ with a generic set of punctures.

\subsection{Maximal gauge group along a tube}\label{section:maximal_gauge_group_along_tube}

In this subsection, we would like to examine the SW curve in the region of the tube when several punctures are brought close together, and to get some preliminary information about the gauge group along the tube, $G_T$. This is a review of a discussion that was done in \cite{Gaiotto:2009we}. Later on we use the characterization of the punctures that are brought together as it appears here. The naive argument that will be given is refined in the consequent subsections.

Suppose then that we bring $m$ regular punctures $P^i$ together. Let them be positioned at $z=\alpha _i$, with $\alpha _i \propto w$ and $w \to 0$ is the limit of bringing them close to each other. As the surface could be of any genus, and there might be additional punctures except for $P^i$, the explicit expression for the entire curve is not straightforward. However, it will be enough to concentrate on the form of the curve in the region of small $z$. This can be written down, and the other details are essentially immaterial for this purpose. The way they do affect the analysis will be explained.
As described in the previous section, the SW differential is $\lambda =xdz$. The curve in the region of small $z$ is approximated by
\begin{equation}
\begin{split}
x^N + \sum _{k=2} ^N \frac{Q_k(z)}{\prod_i (z-\alpha _i)^{p_k^i} } x^{N-k} =0 .
\end{split}
\end{equation}
The polynomials $Q_k$ are determined by the behavior of the Coulomb branch parameters as $w \to 0$. Now let us look at the behavior of the curve in the tube region, $|w| \ll |z| \ll 1$. Substitute $x=y/z$ to get
\begin{equation} \label{eq:SW_curve_around_tube}
y^N + \sum _{k=2} ^N \frac{Q_k(z)}{z^{\sum_i p^i_k-k} } y^{N-k} =0 .
\end{equation}
For some set of punctures $P^i$, we will use the following notation
\begin{equation}
\Delta _k \equiv \sum _i p^i_k - k .
\end{equation}
We always start with $\Delta _2=m-2$. There are essentially two possible qualitative behaviors of $\Delta _k$ for any set of regular punctures. We first review the reasoning for that and then summarize the two types of behavior.

Recall $p_k = k - h(k,P)$, $h(k,P)$ being the row number of the $k$th box in the Young diagram. When $k$ increases by 1, $p_k$ can either increase by 1 if the box is not at the end of a row, or $p_k$ can stay the same if we are at the last box of a row. If a diagram ends with a series of rows of width 1, call this region of the diagram the \textbf{"tip"} of the diagram.

$\Delta _k$ can decrease by 1, stay the same, or increase by $1,2,\dots ,m-1$, as $k$ increases by 1. It decreases by 1 only if we are at the end of a row in each diagram. It stays the same if we are at the end of a row in $m-1$ of the diagrams. 
If $\Delta _k$ decreases by 1 twice consecutively, it means that we were twice at the end of a row consecutively (in all diagrams), and therefore we passed through a row of width 1. Since the row width is non-increasing, we are at the tip of each diagram. All $p_k^i$ will stay the same, and $\Delta _k$ will continue decreasing by 1. We can also note that if $\Delta _k$ decreases by 1 and then stays the same, or stays the same and then decreases by 1, it means that in $m-1$ of the Young diagrams we were at the end of a row twice consecutively, and therefore we are at their tip. We will stay there in these diagrams, and in overall $\Delta _k$ will continue either decreasing or staying the same (it will not increase).

\begin{figure}[h]
\centering
\includegraphics[width=0.6\textwidth]{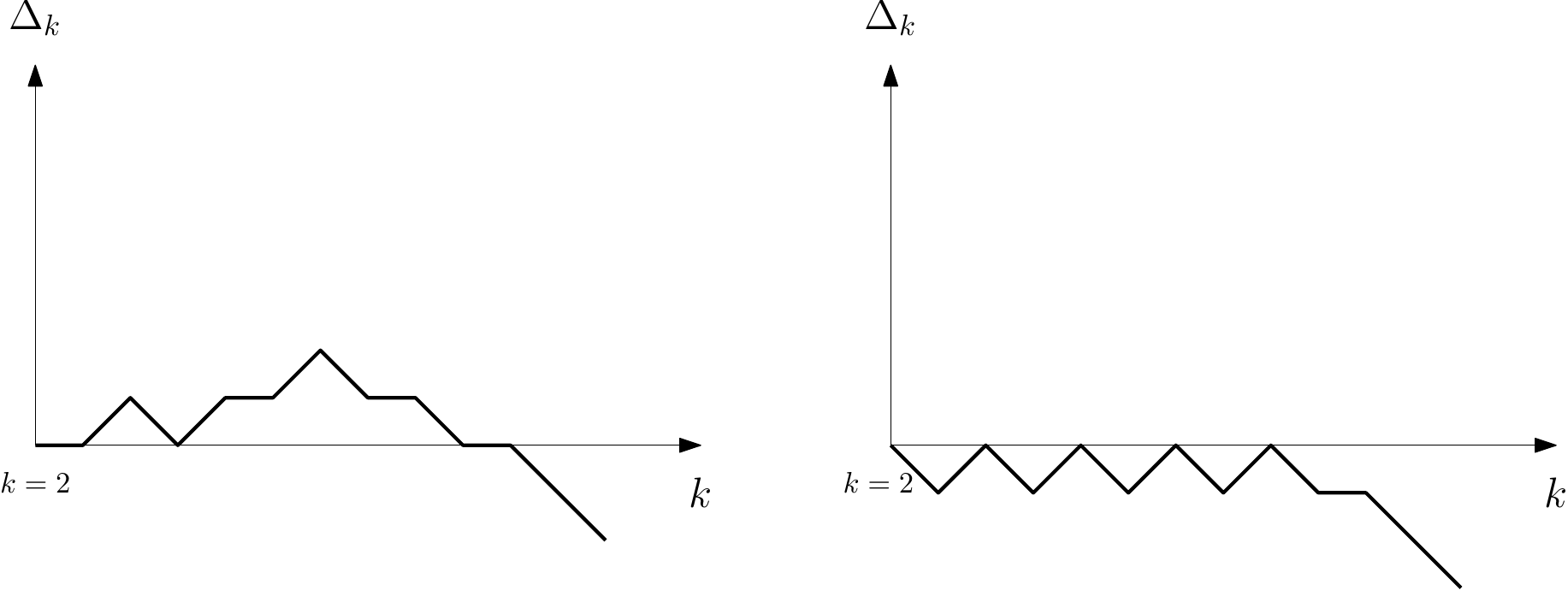}
\caption{The behavior of $\Delta _k$ in the two cases $SU$ and $Sp$.}
\label{fig:Delta_k_behavior}
\end{figure}

What are the options we have? First look at $m=2$ for which we start at $\Delta _2=0$. Suppose first that when going from $k=2$ to $k=3$ we increase at first. Then if we were to go to negative $\Delta _k$, we need to decrease by one from $\Delta =0$, and before that we either stay the same or decrease by 1. In both cases we will not increase anymore, and stay at negative $\Delta $. Therefore in this case, once we crossed to negative $\Delta $ we will not get back. Next suppose $\Delta $ stays the same going from $k=2$ to $k=3$. If we sometime later increase by 1, we are in the previous situation. Otherwise we stayed the same and then decreased, and so again we will stay only at negative $\Delta $, and the behavior is essentially the same as before. Lastly, suppose we first decrease by one going from $k=2$ to $k=3$. This means that $k=2$ is the end of the row for both diagrams and they are of width 2. We then have to increase by 1, decrease by 1, and so on, until one of the diagrams gets to its tip. So after one of the decreases, we will not increase anymore and stay at negative $\Delta $. Qualitatively, we get two options, which are demonstrated in \autoref{fig:Delta_k_behavior}. In the first option, we might not get to a $k$ where $\Delta _k$ becomes negative, and so it happens that $\Delta _k \ge 0$ for all $k$ in this option.

For $m>2$, $\Delta _2=m-2>0$, and we do not have the second behavior of the $m=2$ analysis. To get to negative $\Delta_k $ we must either stay the same and then decrease by 1, or decrease by 1 twice. In both cases we will stay at negative $\Delta _k$. 

Let us call the case in which $\Delta _k$ behaves as in the left diagram of \autoref{fig:Delta_k_behavior} the $SU$ case, and when it behaves as in the right diagram, the $Sp$ case. 
In the SU behavior, $\Delta _k$ is non-negative up to some $k$, and once it crossed the horizontal axis and became negative, it will be non-increasing (as mentioned, it happens that $\Delta _k$ does not get to that region where it is negative). In the Sp case, $\Delta _k$ is zigzagging between $0$ and $-1$ until some stage where it does not increase anymore. It should be kept in mind that the set of punctures $P^i$ is of Sp type exactly when $m=2$ and the first row of each of the two punctures is of width $2$. The rest are of the SU behavior. This is an immediate way to specify whether we are in the SU or Sp case.

\textbf{Define} $\textbf{T}$ to be the last $k$ such that $\Delta _k \ge 0$ for both the $SU$ and $Sp$ cases (and for any $m$). Note that $N$ might be not large enough, in which case the left diagram of \autoref{fig:Delta_k_behavior} may not reach $\Delta _k < 0$. The definition of $T$ will be useful in this situation too.

Now let us apply a naive argument for the gauge group along the tube which will be refined later on. Start with the $SU$ case. The behavior around the tube is governed by \eqref{eq:SW_curve_around_tube}. Choose the Coulomb branch parameters, such that $Q_k(z)$ will behave as $z^{\Delta _k} $ for $k \le T$ around the tube, in the limit $w \to 0$. As will be discussed later, we might not have enough Coulomb branch parameters for that. The terms in \eqref{eq:SW_curve_around_tube} with $k>T$ are negligible. We get an algebraic equation for $y$ with constant solutions. Recall that $\lambda =xdz = y \frac{dz}{z} $ is the SW differential. It follows that these constant solutions give the values of the integrals over the cycles surrounding the tube. These $T$ values (which sum to 0) are vevs of the scalars in the vector multiplet of an $SU(T)$ gauge group. This gives naively $SU(T)$ along the tube.
In the $Sp$ case (for $m=2$), the odd $k$ terms in \eqref{eq:SW_curve_around_tube} are negligible, and we get $USp(T)$ along the tube (the group of rank $T/2$).
If we do not have enough Coulomb branch parameters, we will get a smaller gauge group. In this sense, the gauge group we found given only $P^i$ is the maximal possibility.

\subsection{Decoupling on a sphere using the curve} \label{subsection:decoupling_sphere}

\begin{figure}[h]
\centering
\includegraphics[width=0.5\textwidth]{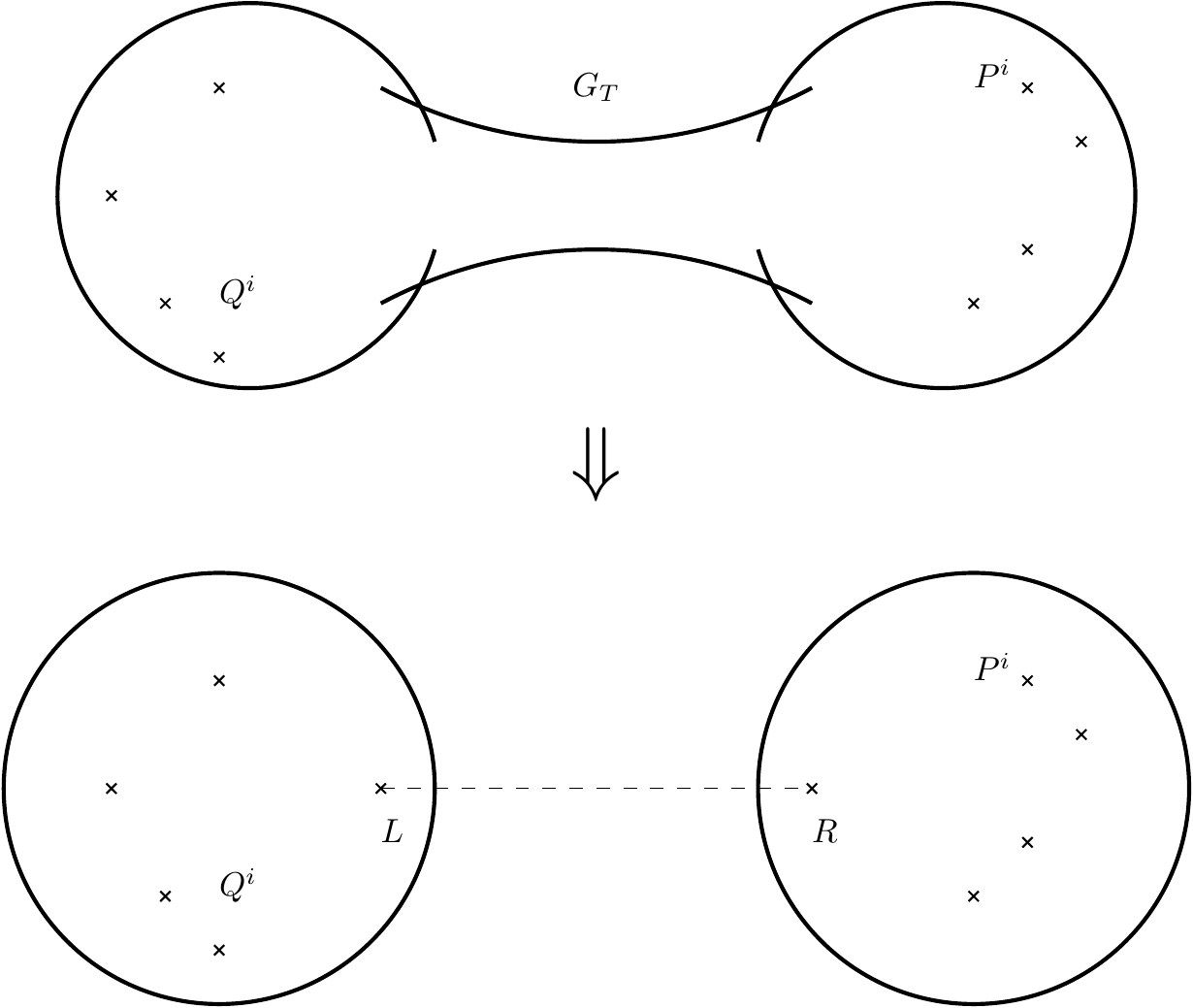}
\caption{Decoupling on a sphere.}
\label{fig:sphere_decoupling}
\end{figure}

The case where $C$ is a sphere is important and instructive. Whenever several punctures on a sphere are close to each other, this is conformally the same as having the rest of the punctures being close to each other. Therefore for a sphere the situation is quite symmetric, in which we have two sets of punctures, see \autoref{fig:sphere_decoupling}. We can think of the punctures on the right as being close to each other, or the punctures on the left being close. As was mentioned before, the curve degenerates into two spheres and a long tube connecting them that represents a weakly coupled gauge group $G_T$. In the simplest description, when the gauge coupling is turned off, we remain with two spheres representing two theories, and at the points where the tube ended before, two new punctures arise, denoted as $L$ on the left sphere and $R$ on the right one. The question we ask is what are the resulting two theories and what is $G_T$ in general.

In the analysis of the previous subsection, we considered only the side of the surface with the punctures brought close to each other. We did not have enough information to determine what is the resulting $G_T$. Indeed, it cannot be fixed uniquely by considering punctures on one side alone, as will be discussed in \autoref{sec:decoupling_punctures_fix_tube}.

This is related to what was done in \cite{Chacaltana:2012ch}. There, at first stage each side in the degenerating sphere is considered separately. The analysis of the SW curve of the previous subsection is the same as that in section 3 of \cite{Chacaltana:2012ch}\footnote{It is applied there to theories beyond the $A_{N-1} $ which we consider here, and also in section 2.2 of \cite{Chacaltana:2013oka} and section 2.4.5 of \cite{Chacaltana:2014jba}.}. For each of the two sides the gauge group on the forming tube was found, as well as the puncture that would be created on the other side of each tube. However these two gauge groups are what we referred to as the maximal gauge groups. In order to get the final result of the decoupling, some sewing procedure must be done. One approach to do that which was used in \cite{Chacaltana:2012ch} is to consider the sphere containing the two new punctures that were found and to use the description of this sphere in \cite{Gaiotto:2011xs} as a supersymmetric non-linear sigma model. In this description the product of the two maximal gauge groups is Higgsed to the actual gauge group $G_T$ that eventually emerges. Additionally, it is important that the leftover spheres in this description may also change by Higgsing coming from the D-terms and the F-terms, and this should be taken into account. In the following we will use a different procedure and find directly the result of the sewing.

In the case of a sphere, we can write simply the full curve. We will perform an analysis which is more precise than that of the previous subsection, and at the end of this subsection we summarize the result that we get. The main question in such an analysis is how the Coulomb branch parameters should behave as a function of the scale of the punctures that are taken close to each other (as $w \to 0$ above), with the requirement that the curves that will be left on both sides make sense. This question can also be rephrased as how the zeroes of the meromorphic differentials $\phi _k$ should behave in the limit we take.

Let us then write the curve of the theory.
Denote the set of punctures on the right by $P^i$, and associate to them $\Delta _k^R = \sum _i p^i_k - k$ as before; similarly for the left side we define $\Delta ^L_k=\sum_i q_k^i-k$ with $Q^i$ the punctures on the left. The curve is of the form
\begin{equation} \label{eq:sphere_curve}
\begin{split}
& \lambda ^N+ \phi _2\lambda ^{N-2} + \dots + \phi _N=0, \qquad \lambda =x\, dz \\
& \phi _k=\frac{Q_k(z)}{\prod (z-\alpha _i)^{p^i_k} \prod (z-\beta _i)^{q^i_k} } dz^k .
\end{split}
\end{equation}

Suppose we bring the punctures on the right side close together and to the point labelled $z=0$. The position of these points is proportional to some $w$ and we take $w \to 0$. The $\phi_k$'s appearing in the curve in this limit $w \to 0$ are
\begin{equation} \label{eq:approx_phi_k}
\phi_k \sim \frac{Q_k(z)}{z^{\Delta ^R_k+k} } dz^k = u_k \frac{(z-z^{(k)} _1) \dots (z-z^{(k)} _{n_k})}{z^{\Delta ^R_k+k} } dz^k .
\end{equation}
When we take $w\to 0$ we have to decide how the parameters behave in order to get a sensible curve which will remain on the LHS. The degrees of the poles in \eqref{eq:sphere_curve} fix the degree of the polynomial in the numerator of $\phi _k$, 
\begin{equation}
n_k=\Delta _k^L+\Delta _k^R
\end{equation}
where $n_k$ is defined in \eqref{eq:approx_phi_k} (it is related to $d_k$ from the previous section by $d_k=n_k+1$).

\begin{figure}[h]
\centering
\includegraphics[width=0.6\textwidth]{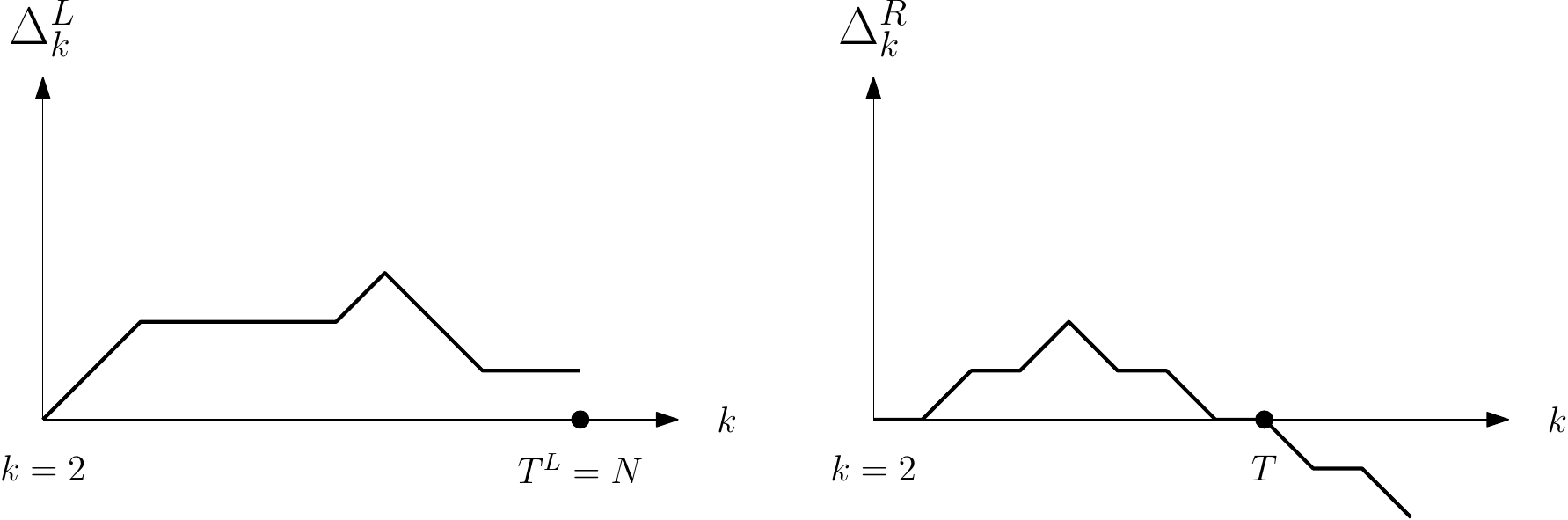}
\caption{Both $\Delta _k^L$ and $\Delta ^R_k$ are of the $SU$ case which means that up to some $T$, $\Delta _k \ge 0$ and after it $\Delta _k<0$. Here we have $T^L=N$ ($N$ is the last value $k$ obtains), while $T^R$ may be smaller or equal to $N$. The actual values in the plot are chosen arbitrarily and should not have to correspond to a realistic case, but are shown for demonstration purposes only.}
\label{fig:two_spheres_basic_behavior}
\end{figure}

For a set of punctures we defined $T$ as the last $k$ such that $\Delta _k \ge 0$.
Here we have such $T$ associated to the left punctures and the right punctures, $T^L$ and $T^R$.\\
We start by assuming that both $\Delta _k^L$ and $\Delta _k^R$ are of the $SU$ case. We cannot have simultaneously $T^R<N$ and $T^L<N$ because then $n_N<-1$. Then suppose $T^L=N$, $T^R=T$; we are in the situation depicted in \autoref{fig:two_spheres_basic_behavior}.

The following discussion is done for $\Delta _k^R$, but will be used afterwards for $\Delta ^L_k$ and $\Delta^R _k$ interchanged. Statements as "if $\Delta ^L_k \ge 0$ then... " which seem redundant for $\Delta ^L_k$, are not redundant when we switch $\Delta^L_k$ by $\Delta ^R_k$. The following is then a general discussion. \\
Fix some $k$. If $\Delta ^R_k\ge 0$ we ask if we can "flatten" $\Delta ^R_k$, by which we mean that we can have enough $z^{(k)} _i$'s in $Q_k$ that we can take to 0 together with $w\to 0$, such that the power of $1/z$ in $\phi_k$ will reduce from $\Delta^R _k+k$ to $k$. This can be done only when $n_k\ge \Delta ^R_k$, which is $\Delta ^L_k \ge 0$. So we will say that we can "flatten" $\Delta ^R_k$, if $\Delta ^L_k \ge 0$.
\begin{itemize}
\item If we are not able to flatten $\Delta ^R_k$, we know we cannot be left with a pole of order $> k$ (since we do not have those in the superconformal theories we consider), and we are forced to take $u_k=0$ ($u_k$ defined in \eqref{eq:approx_phi_k}). 
\item If we can flatten $\Delta ^R_k$, we are brought to $(A+Bz+Cz^2+\dots)/z^k$ in $\phi_k$. Note that $z=0$ becomes the position of the created puncture $L$ on the remaining LHS sphere. 

The constants $A$ in the various $A/z^k$ of $\phi_k$ are exactly the Coulomb branch parameters of the gauge group along the tube. This is so, because these terms give in the curve $x^N+ \sum _{k=2} ^T \frac{A_k}{z^k} x^{N-k}=0$, which after a change of variables $x=y/z$ gives an algebraic equation for $y$: $y^N+\sum _{k=2} ^T A_k y^{N-k} =0$. As explained in the previous subsection, the solutions for this equation give the Coulomb branch parameters of the gauge group (denoted usually by $a_i$).

Since we are interested in the LHS sphere, we take the $A$'s to 0, which just amounts to taking the Coulomb branch parameters of the gauge group along the tube to 0. This does not affect $B,C,\dots $ . Therefore we get that when $\Delta ^R_k\ge 0$ can be flattened, $l_k=k-1$.
\item If $\Delta ^R_k<0$, we do not have to do anything and therefore $l_k=k+\Delta ^R_k = \sum_i p_k^i$. \footnote{We could think that a pole of lower order can be obtained as well. This is addressed in the discussion about surfaces of general genus (\autoref{sec:appendix_g_ge1_analysis}).}
\end{itemize}

Note that in the second case, it might be that we do not have the $B,C,\dots $ terms. When we flattened $\Delta ^R_k \ge 0$, we had to use $\Delta ^R_k$ number of the $z^{(k)} _i$'s in \eqref{eq:approx_phi_k}. There will be no $B,C, \dots $ terms exactly when $n_k=\Delta ^R_k$, or equivalently $\Delta _k^L=0$.  In this case, even though we can flatten $\Delta ^R_k$, we will be left with $A/z^k$ and then take $A \to 0$ to get the massless curve. Therefore we will get $\phi ^L_k=0$ in the SW curve of the LHS sphere for these values of $k$, and $l_k$ is not defined by the SW curve. However, the assignment of $l_k=k-1$ above can still be used as will be explained in a moment. The idea briefly is that  if we have on the LHS sphere the punctures $q^i_k$ and $l_k=k-1$ then the sum $\sum _i q^i_k+l_k=\Delta _k^L +k+k-1=2k-1$ and there is no $k$-differential with poles of total order $2k-1$ on the sphere, forcing indeed $\phi _k^L=0$.

Let us apply the above to the situation of \autoref{fig:two_spheres_basic_behavior}.
For $k\le T$, we can "flatten" all the $\Delta ^R_k$ since $\Delta ^L_k \ge 0$, and therefore necessarily $l_k=k-1$ there. For  $k>T$, $l_k=\sum_i p_k^i$. \\
The same can be done in the other direction, by asking what is left on the right hand sphere. For this we just interchange the roles of $\Delta ^R_k$ and $\Delta ^L_k$. $\Delta ^L_k$ is always non-negative, but we cannot flatten all the $\Delta ^L_k$ --- we see this by $\Delta ^R_k$. We can flatten only $k\le T$. Therefore $r_k=k-1$ for $k\le T,$ where $r_k$ is the puncture created on the RHS sphere. For $k>T$, we cannot flatten $\Delta ^L_k$, so we will get $\phi_k^R=0$ in the SW curve of the RHS sphere. $r_k$ is not defined by the SW curve for $k$ in this range. \\
We had one Coulomb branch parameter (an A) for each $2 \le k \le T$. From the point of view of $\Delta ^R_k$ this was so because $\Delta ^R_k \ge 0$ for these $k$'s and all of them can be flattened because $\Delta ^L_k \ge 0$. From the point of view of $\Delta ^L_k$ this was so because even though $\Delta ^L_k \ge 0$ for all $k$'s, only for $2 \le k \le T$, $\Delta ^L_k$ can be flattened because only there $\Delta^R _k \ge 0$. We then have along the tube $SU(T)$.

We have seen several cases in which we get $\phi _k=0$ in the LHS or RHS spheres (and therefore we cannot get the pole structures of $L$ or $R$). There are basically two ways to approach this. We will start with the first approach by describing an additional meaning in which we can assign a value to $l_k$ or $r_k$ in such situations.

In general, define $d _k$ associated to some theory to be the number of Coulomb branch parameters of dimension $k$. We had a formula \eqref{eq:Coulomb_branch_graded_dimension} for $d_k$ in terms of the pole structures for a non-zero $\phi _k$. The theory we started from had $d_k=n_k+1$ parameters of dimension $k$. For $k \le T$ one parameter became the single Coulomb branch parameter of dimension $k$ for the gauge group along the tube (the one that was denoted by $A$). As we said before, after flattening $\Delta ^R_k$, we will be left with $(A+Bz+...+ Dz^{n_k-\Delta ^R_k} )/z^k$ and take $A \to 0$. This leaves $n_k-\Delta ^R_k=\Delta^L_k$ parameters on the LHS sphere. Therefore $d_k^L=\Delta ^L_k$ and similarly $d_k^R=\Delta ^R_k$. As expected, the number of parameters is conserved $d_k=d_k^L+d_k^R+d_k^{\text{tube}}=\Delta ^L_k+\Delta ^R_k+1$. For $k>T$, no parameters go to the tube. One of the remaining curves' $\phi _k$ is 0 as we saw: in \autoref{fig:two_spheres_basic_behavior} it was $\phi _k^R$, so $d_k^R=0$. $\phi _k^L$ was just inherited from $\phi _k$, and therefore $d_k^L=d_k$. So again the number of parameters is conserved, $d_k=d_k^L+d_k^R+d_k^{\text{tube}} $.

Now take some $k$ for which we assume that $\Delta ^R_k=n_k>0$ (and then $\Delta ^L_k=0$), corresponding to the first situation of a vanishing $\phi _k$ that we encountered. According to the discussion above, $\phi _k^R$ is non-zero, but $\phi _k^L= d_k^L=0$. Now suppose, by definition, that we want to maintain the equation
\begin{equation}\label{eq:graded_dimension}
d_k=\sum p_k^i-2k+1 \qquad (p_k^i \text{ here denote general punctures})
\end{equation}
for the LHS sphere, even though for those $k$'s $\phi _k^L=0$ and $l_k$ is not defined by the SW curve. This will imply that $l_k=d_k^L-\sum q_k^i+2k-1=k-1-\Delta ^L_k=k-1$. In this sense, we get again $l_k=k-1$. For $k$'s satisfying $\Delta ^L_k=n_k$ we get by the same reasoning that even though $r_k$ is not defined by the SW curve, to preserve \eqref{eq:graded_dimension} we still can define $r_k=k-1$ as we did before naively.

For $k>T$ we said that $r_k$ is not defined by the SW curve. Using the definition we just had, we can similarly define $r_k$ there. $d_k^R=0$, so $r_k=d_k^R-\sum p_k^i+2k-1=2k-1-\sum p_k^i$. These $r_k$'s in the range $k>T$ satisfy $r_k \ge k$. Punctures having a pole structure greater than $k-1$ are called \textbf{irregular punctures}. Irregular punctures and this approach of using the graded dimension of the Coulomb branch are discussed in \cite{Chacaltana:2010ks}.

Note that we could be in the situation in which we started from a curve on $C$ in which $\phi _k=0$ for some $k$. The pole structures of the punctures on $C$ ($p^i_k$ and $q^i_k$) are not defined by the massless curve, but are given as part of the definition through a Riemann surface with co-dimension 2 defect operators. We can do the same procedure: the resulting curves will have of course $\phi _k^L=\phi _k^R=0$ for the $k$'s with $\phi _k=0$, and then $d_k^L=d_k^R=0$ again fix the $l_k$ and $r_k$ from $q^i_k$ and $p^i_k$.

We would like now to complete the discussion of all the possible decouplings of a sphere into two spheres in the current approach, and after this to describe another equivalent description in which we do not use irregular punctures. If both $\Delta _k^L$ and $\Delta _k^R$ are of the $SU$ case, as was discussed $T^L<N$ together with $T^R<N$ is not possible because it implies $d_N<0$. Therefore if both $\Delta _k$ are of the $SU$ case, we have one with $T=N$, say $T^L=N$ and the other $T^R=T \le N$. Both $\Delta _k^L$, $\Delta _k^R$ cannot be of the $Sp$ case because they give once again some $d_k<0$ \footnote{$\Delta _k<0$ for all $k>2$ satisfies both the $SU$ and $Sp$ behaviors, and we consider it to be of the $SU$ type.}. If one of the $\Delta _k$'s is of the $Sp$ type, then the other must be an $SU$. If the $Sp$ have $T<N$, the $SU$ must have $T=N$ (otherwise $d_N<0$). If on the other hand the $Sp$ have $T=N$, the $SU$ can have $T=N$ or $T<N$. If it is $T<N$, it actually must be $T=N-1$ because otherwise $d_{N-1} <0$. The following three cases then cover all the situations
\begin{enumerate}
\item Left : $SU$, $T^L=N$. Right : $SU$, $T^R=T \le N$.
\item Left : SU, $T^L=N$. Right : $Sp$, $T^R =T \le N$.
\item Left : $Sp$, $T^L=N$. Right : $SU$, $T^R=T=N-1$. Possible only for even $N$.
\end{enumerate}
(we have chosen what is left and what is right for convenience). \\
We described how $l_k$ and $r_k$ are fixed in general. Let us summarize it for instance for $l_k$. $r_k$ is obtained in the same way by switching left and right. Consider $\Delta _k^R$ and suppose $\Delta _k^R \ge 0$. If $\Delta_k^L > 0$, $\Delta _k^R$ can be flattened and we are left with a non-zero $\phi _k^L$. In this case $l_k=k-1$. If $\Delta _k^L  = 0$ we can flatten $\Delta _k^R$ but are left with $\phi _k^L=0$ when taking $A=0$. If $\Delta _k^L<0$ we cannot flatten $\Delta _k^R$, and must take $u_k=0$ to get $\phi _k^L=0$. Therefore for $\Delta _k^L \le 0$, $\phi _k^L=0$, or equivalently $d_k^L=0$. We now use \eqref{eq:graded_dimension} to fix $l_k= d_k^L-\sum_i q_k^i +2k-1=k-1 - \Delta _k^L$. In case where $\Delta _k^R<0$ we do not need to do anything and just have $l_k= \sum_i p_k^i$. \\
These considerations assume $\phi _k \neq 0$, but hold also in case that $\phi _k=0$. To see this, note first that clearly $d_k^L=d_k^R=0$ for such $k$. If $\Delta _k^R \ge 0$ then $\Delta _k^L=n_k-\Delta _k^R=-1-\Delta _k^R<0$. $l_k$ is fixed by $l_k=d_k^L-\sum _i q_k^i+2k-1=k-1-\Delta _k^L$ as obtained in the previous paragraph. If $\Delta _k^R<0$, again $\Delta _k^L = -1-\Delta _k^R$. Still $l_k=-\sum _i q_k^i +2k-1=k-1-\Delta _k^L = k+\Delta_k^R =\sum _i p_k^i$ as before.\\ 
To summarize,
\begin{equation}
l_k = \begin{cases}
k-1 & \Delta _k^R \ge 0, \Delta _k^L \ge 0 \\
k-1-\Delta _k^L & \Delta _k^R \ge 0, \Delta _k^L<0 \\
\sum _i p_k^i & \Delta _k^R<0
\end{cases} .
\end{equation}
This equation and the one corresponding to switching left and right fix $l_k$ and $r_k$ in the three cases mentioned above.

\begin{figure}[h]
\centering
\includegraphics[width=0.8\textwidth]{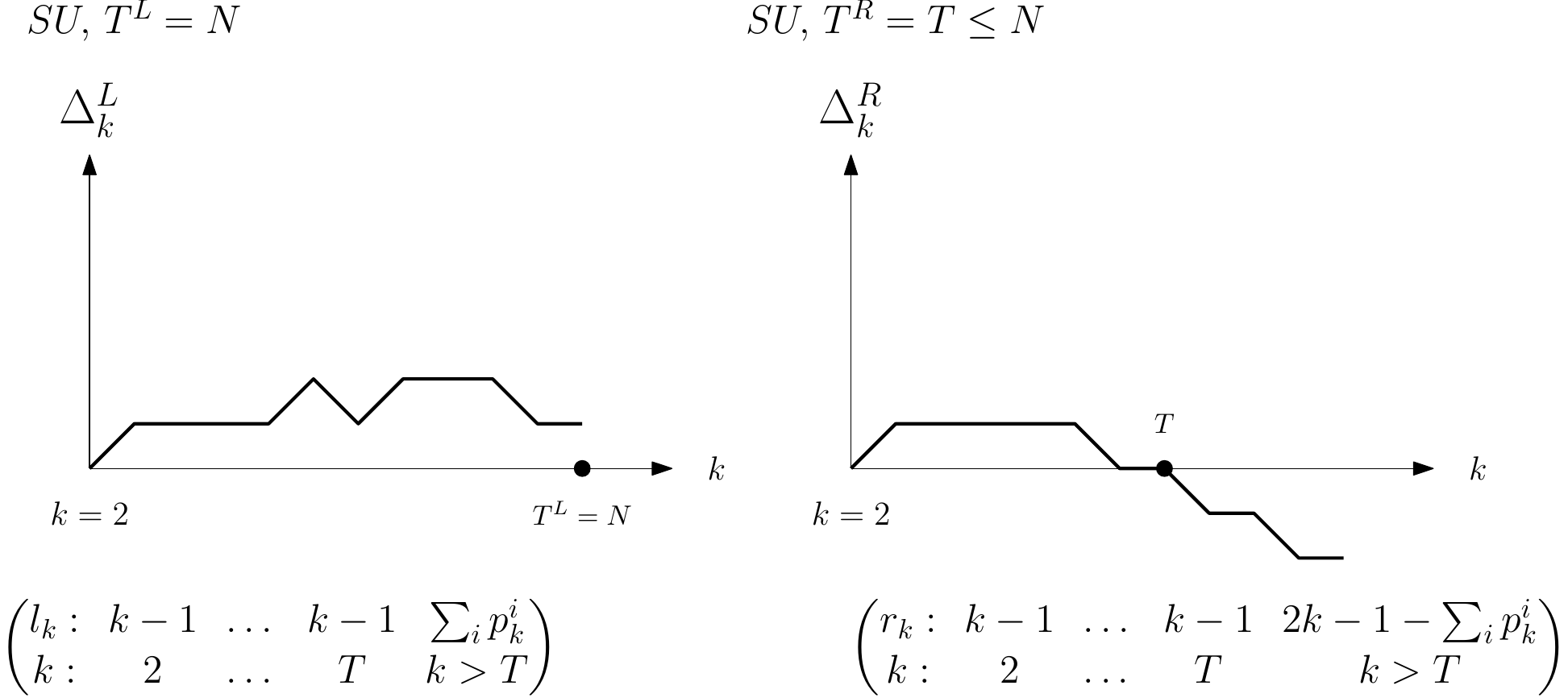}
\caption{First case. We get in the tube $SU(T)$. If $T=N$ both $l_k$ and $r_k$ are full punctures.}
\label{fig:two_spheres_SUSU}
\end{figure}

\begin{figure}[h]
\centering
\includegraphics[width=0.8\textwidth]{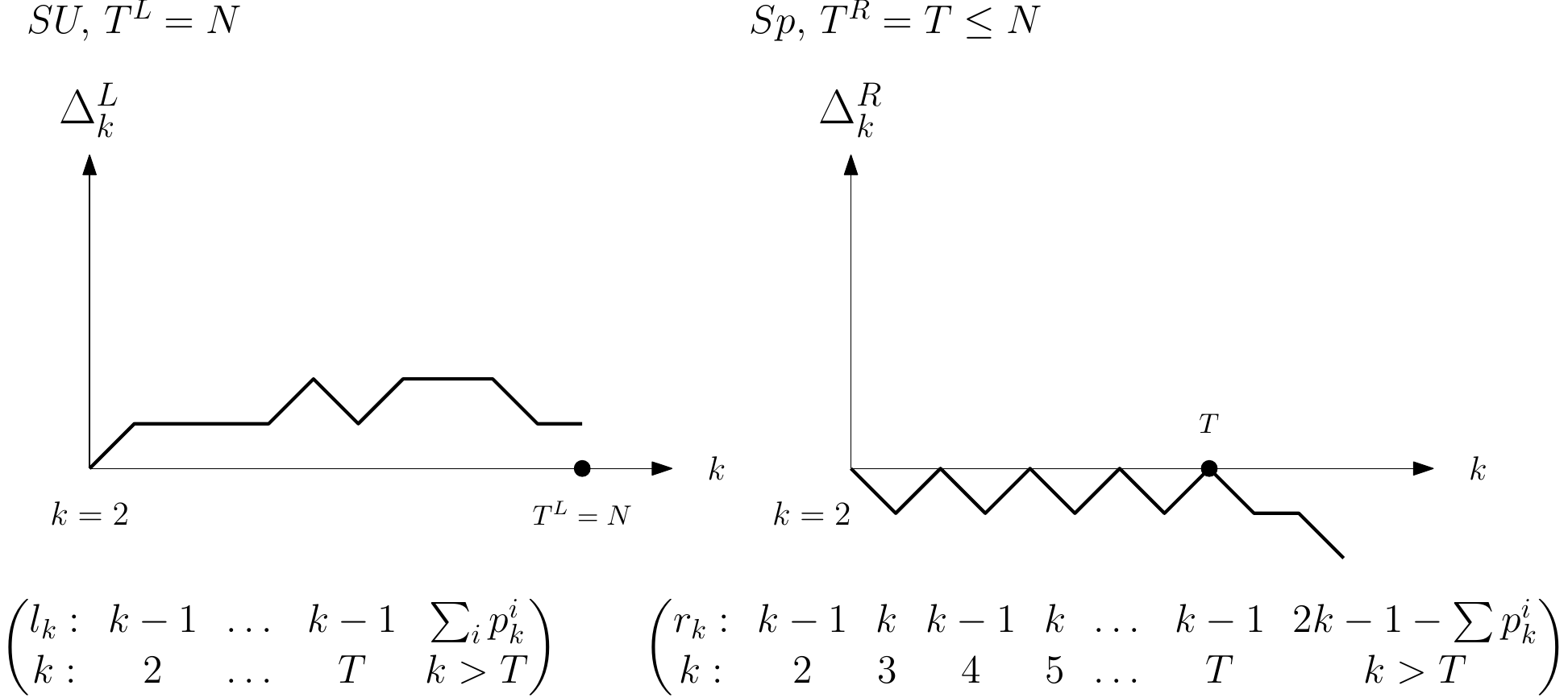}
\caption{Second case. We get in the tube $USp(T)$.}
\label{fig:two_spheres_SUSp1}
\end{figure}

\begin{figure}[h]
\centering
\includegraphics[width=0.8\textwidth]{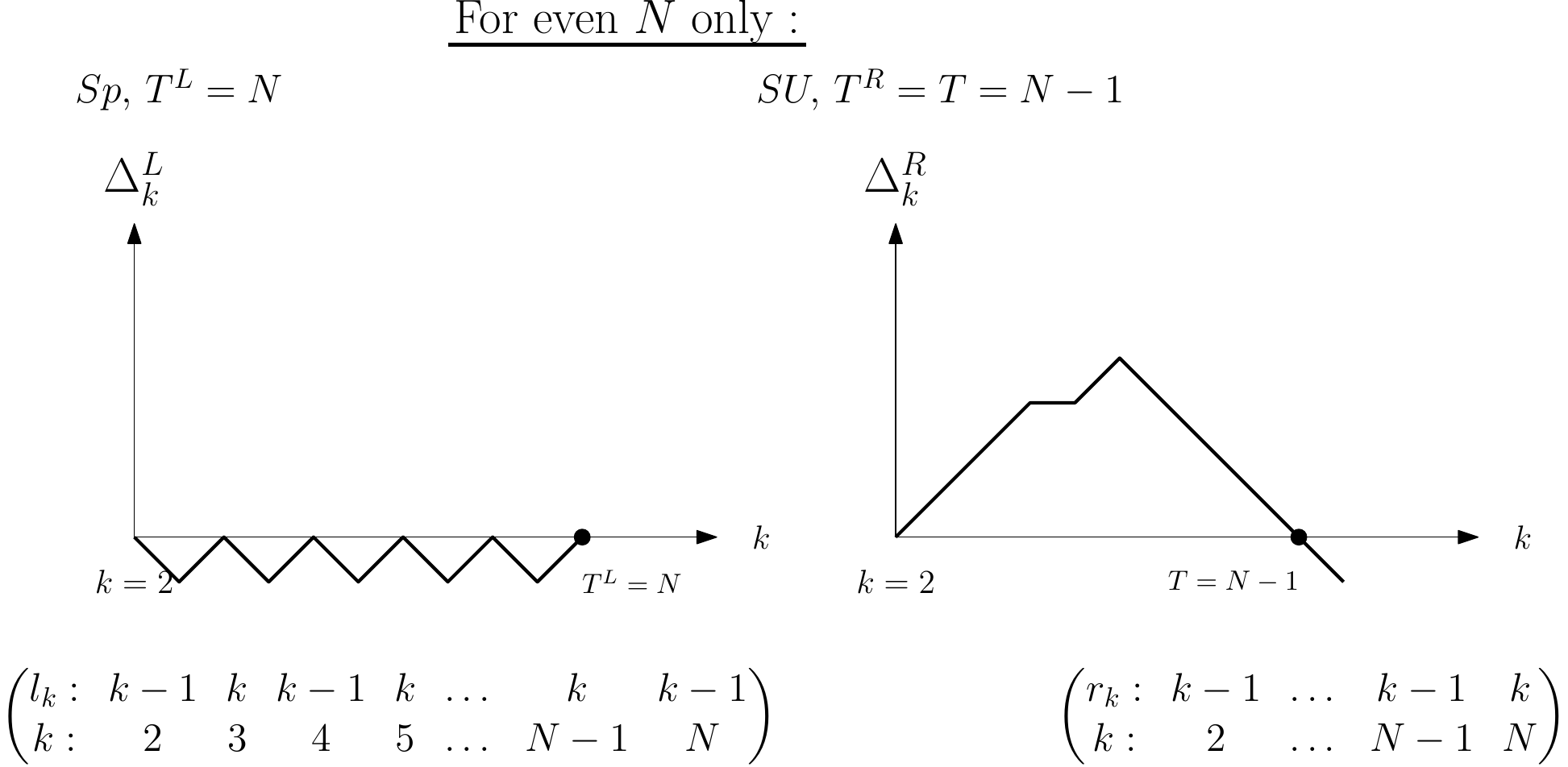}
\caption{Third case. Possible only for even $N$. In the tube we get $USp(N-2)$.}
\label{fig:two_spheres_SUSp2}
\end{figure}

Every given situation falls into one of these three cases, which are shown in figures \ref{fig:two_spheres_SUSU},\ref{fig:two_spheres_SUSp1},\ref{fig:two_spheres_SUSp2}. The plots are for illustration, and the essential behavior is indicated above the plots. As we saw, there is one Coulomb branch parameter of dimension $k$ in the gauge group along the tube exactly for $\Delta _k^L,\Delta _k^R \ge 0$. This gives us the gauge group along the tube in the three cases: $SU(T)$, $USp(T)$ and $USp(N-2)$. After the decoupling the resulting two theories are those defined by a sphere with the punctures $P^i$ and $R$ for the RHS theory, and $Q^i$ and $L$ for the LHS theory. $L$ and $R$ are indicated in the appropriate figure.

The irregular punctures are $R$ in the first case if $T<N$, $R$ in the second case, and both $L$ and $R$ in the third case. We have two irregular punctures at both ends of the tube only for even $N$, in which case we get $USp(N-2)$ along the tube.
The only case in which both punctures at the ends of the tube are regular is \autoref{fig:two_spheres_SUSU} with $T=N$, in which both punctures are full punctures ($l_k=r_k=k-1$ for all $k$) and the gauge group along the tube is $SU(N)$.

We can describe the result of the decoupling using only the familiar regular punctures. In all of the resulting spheres that we have found having an irregular puncture, the $\phi _k$ are zero starting from some $k$ \footnote{This might also happen in theories with regular punctures only.}. There are then a set of branches of the curve that are decoupled from the rest of it. This is the same as
\begin{itemize}
\item A usual $A_{N'-1} $ theory with lower $N'<N$ having only regular punctures, with the curve obtained by cancelling a common factor of $x$ in \eqref{eq:general_curve_form_class_S} from the curve with the vanishing $\phi _k$'s.
If all $\phi _k=0$ for $k \ge 2$ we have only the second ingredient below.
The punctures are obtained by truncating the pole structures to $2 \le k \le N'$.
\item Plus possibly additional free hypermultiplets.
\end{itemize}

The number of additional free hypermultiplets in such a theory can be found as follows. We started from some theory $C$ and after a decoupling limit, had a theory $C_1$, a theory $C_2$ which is a theory of regular punctures, and $n_h$ additional free hypermultiplets. Each of $C$, $C_1$ and $C_2$ is a theory of regular punctures. For each theory, a number of effective hypermultiplets and vector multiplets was defined, by a relation to the $a,c$ anomalies \cite{Gaiotto:2009gz}. 
Subtracting the number of effective hypers of $C_1$ and $C_2$ from $C$ we find $n_h$. We will not quote this simple algebraic calculation, but merely give the result.

In the Sp behavior, the resulting curve on the Sp side will be trivial ($x ^N=0$), and the theory there is then only a set of free hypermultiplets.

The missing ingredient is an expression for $N'$ in the SU case. We will use results that will be obtained later on in order to express the different quantities using the diagrams of the punctures. According to what we saw, in the resulting theories of interest, the last $k$ for which $\phi _k \neq 0$ and afterwards all $\phi _k=0$ is the last $k$ such that $\Delta _k>0$. Using the discussion in \autoref{section:maximal_gauge_group_along_tube} and equation \eqref{eq:T_diagrammatic}, if there exists a last $k$ such that $\Delta _k>0$ then it is given by $\sum _{i=1} ^{\alpha -1} P_i^1$ (we use the conventions and notations from the text around \eqref{eq:T_diagrammatic} and will write these below when we summarize the result \footnote{In these conventions we choose $P^1$ to be of largest $p$. Note that there might be several punctures with the largest $p$. In all the calculations (such as that of $T$ in \eqref{eq:T_diagrammatic}) it does not mater which one we choose, except for calculating $N'$ in some cases where we have $m=2$ punctures. In these cases, if only one of them has $P_{\alpha } >1$ we choose it as $P^1$ (and otherwise it does not matter which one is chosen).   }).

If there is no $\Delta _k>0$, the resulting theory is a theory of free hypermultiplets ($N'=0$). By following the lines of the analysis of the behavior in the SU case, we can construct quite easily the diagrams that will result in all $\Delta _k \le 0$ by following the behavior of $\Delta _k$ as we append each box. Since $\Delta _2>0$ if $m>2$, this can happen only with two punctures $P^1,P^2$ that are brought together to form the decoupling sphere in which we are interested. By doing this exercise, we find that there is no $\Delta _k>0$ in three basic cases (which are not distinct in the way they are written). 
In the first option we have a simple puncture and some other puncture $P^1$. In this case $\alpha =1$ and the resulting theory is only a theory of $P^1_1(P^1_1-P^1_2)$ free hypermultiplets. The second option is when we have $P^2_1=2$, $P^1_1=3$, $P^1_2 \le 2$. Finally, we can have any $P^1_1 \le 3$ and $P^2$ of rows $2,2,1,1,\dots $.  \\
For these cases we just need the number of free hypers. In general, as was explained above, the number of effective hypers minus the number of effective vector multiplets can be calculated for the decoupled RHS theory in the first general case (\autoref{fig:two_spheres_SUSU}). The diagrammatic description which will be given later on is used in this calculation. The number that is found is (still ordering $p^1 \ge p^i$)
\begin{equation}
\begin{split}
& n_{\Delta } =n_h-n_v=-1+ \frac{1}{2} P^1_{1,\alpha } (P^1_{1,\alpha } -P^1_{\alpha +1} ) +\frac{1}{2} \sum _{i=1}^{\alpha} P^1_{1,i} (P^1_i-P^1_{i+1} ) + \sum _{j \ge 2} n_{\Delta } (P^j) \\
& n_{\Delta } (P) =- \frac{N}{2} +\sum _i \frac{1}{2} P_{1,i} (P_i-P_{i+1} )  \ge 0 , \qquad 
P_{1,i}  \equiv \sum _{j=1} ^i P_j .
\end{split}
\end{equation}
In these cases in which there is no $\Delta _k>0$ we find that the total number of free hypers is
\begin{equation} \label{eq:free_hypers_SU_RHS_when_delta_k_le_0}
n_h=-1+ \frac{1}{2} P^1_{1,\alpha } (P^1_{1,\alpha } -P^1_{\alpha +1} ) +\frac{1}{2} \sum _{i=1}^{\alpha} P^1_{1,i} (P^1_i-P^1_{i+1} ) + \sum _{j \ge 2} n_{\Delta } (P^j) .
\end{equation}

Let us summarize how the result of decoupling can be found very easily, in terms of usual theories with regular punctures. We are given a sphere $C$ with a set of punctures $P^i$ on the right and a set of punctures $Q^i$ on the left. We will assume that they are ordered such that $p^1 \ge p^i$ for any $i$ (see the previous section for the definition of $p$), and the same for $Q^1$ relative to $Q^i$ \footnote{For choosing $P^1$ among several punctures with the largest $p$, see the previous footnote.}.
 Up to a possible renaming of what we call left and right, any given configuration should fall into one of the following cases, where by "$P^i$ are of Sp type" we simply mean that we have 2 punctures with their first row being of width 2 and all the rest is referred to as SU type:
\begin{enumerate}
\item $P^i$ and $Q^i$ are of SU type, and $\sum_i q^i \ge N$. \\
In that case, we have along the tube $SU(T)$ with $T=\sum _{i=1} ^{\alpha} P^1_i$ where we denote $\alpha =\sum _{j \ge 2} p^j$. The resulting theory on the LHS is given by an $A_{N-1} $ theory with the punctures $Q^i$ and additionally $l_k=\min(k-1,\sum _i p_k^i)$. We will show that the diagram of $L$ is the diagram of $P^1$ with the $\alpha +1$ first rows merged to a single one.\\
The theory on the RHS is a theory in $A_{N'-1} $ where $N' = \sum _{i=1} ^{\alpha-1} P^1_i$, with the punctures $P^i$ (truncated to $k=2,\dots, N'$) and an additional full puncture instead of the tube, plus $(N'+P^1_{\alpha } )(P^1_{\alpha } -P^1_{\alpha +1} )$ free hypers. \\
If there are only two $P^i$ punctures such that one of them is a simple puncture, or one of them has $P_1=2$ and the other $P_1=3$, $P_2 \le 2$, or one of them has $P_1 \le 3$ and the other is of rows $2,2,1,1,\dots $, then the RHS is a free theory with the number of hypers given by \eqref{eq:free_hypers_SU_RHS_when_delta_k_le_0}.
\item $Q^i$ are of SU type while $P^i$ are of Sp type, and $\sum _i q^i \ge N$. \\
Along the tube we get $USp(T)$ where now $\alpha = P^2$ since there are only 2 punctures on the RHS. The resulting theory on the LHS is described just as in the first case. The theory on the RHS is just $2\alpha (2-P^1_{\alpha +1}) $ free hypers.
\item The third case is quite special. It occurs only for even $N$. On the LHS there are 2 punctures with all rows being of width 2. On the RHS $P^i$ are of the SU type with $\sum _i p^i = N-1$. \\
On the tube we have $USp(N-2)$. The LHS theory is a free theory of $2N$ hypers. The theory on the RHS is as described in the first case.

\end{enumerate}

What is needed above are only the rows structure of the Young diagrams of the punctures, and this specification is therefore very easy to apply. \footnote{Note that we assume that the sphere we began with is a legitimate theory, since not any collection of punctures on the Riemann sphere is an acceptable theory.}

\subsection{$g \ge 1$ surfaces} \label{subsection:decoupling_g_ge1}

\begin{figure}[h]
\centering
\includegraphics[width=0.6\textwidth]{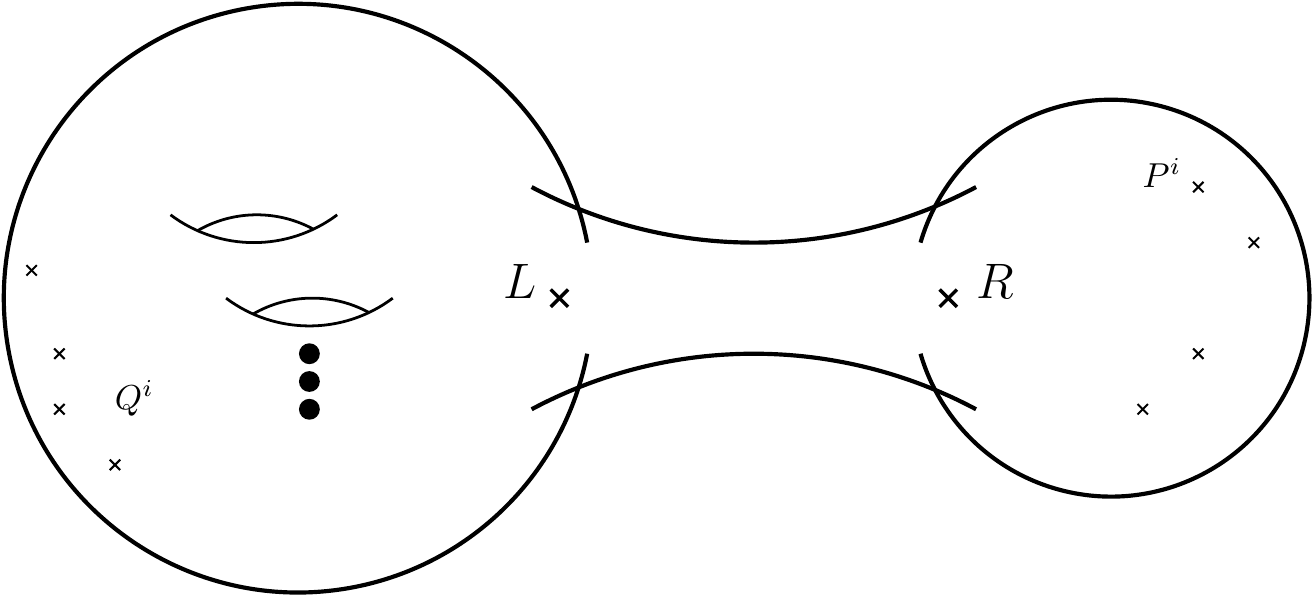}
\caption{Decoupling in a $g \ge 1$ surface. The punctures $L$ and $R$ are shown but do not appear until the complete decoupling of the tube. }
\label{fig:g_ge1_decoupling}
\end{figure}

Suppose we have a $g \ge 1$ surface with regular punctures, and we bring several of them close together. We would like to get the resulting theories and tube as was done for the sphere. The scenario we consider is depicted in \autoref{fig:g_ge1_decoupling}.

In the case of the sphere, the situation between the two sets of punctures was symmetric. We could learn on each side by bringing the punctures on the other side close to each other. In a $g \ge 1$ surface the situation is not symmetric. We bring punctures say on the right close to each other. A sphere bubbles off and a long tube is formed, connecting the sphere and the remaining surface (as in \autoref{fig:g_ge1_decoupling}). An additional complication is that on the sphere we could use the full curve, which is more involved for a general surface. We should avoid writing its full form, and instead concentrate on the region of the punctures that are brought together, and be able to get both of the resulting theories. 

This analysis appears in \autoref{sec:appendix_g_ge1_analysis}.
The answer is that any situation will be either the first or the second scenario described for the sphere (without the third one). The gauge groups along the tube and the resulting two theories are the ones described there. In the convention used, the RHS theory is a sphere, while the LHS is of the same genus as that of $C$.

\subsection{Do the decoupling punctures fix the tube?} \label{sec:decoupling_punctures_fix_tube}

\begin{figure}[h]
\centering
\includegraphics[width=0.6\textwidth]{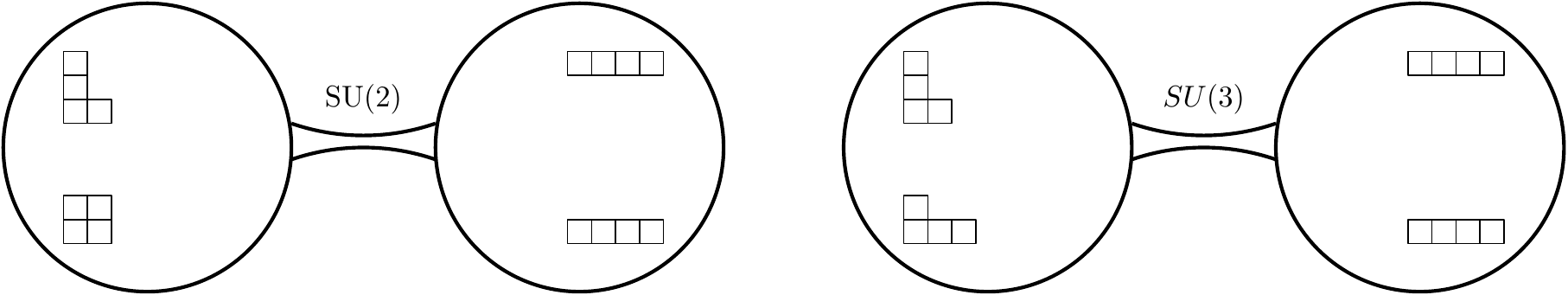}
\caption{Several options for decoupling of the same right sphere (see \cite{Chacaltana:2010ks},\cite{Chacaltana:2011thesis} for many examples). }
\label{fig:several_decoupling_options}
\end{figure}

In this section, we saw that given the punctures on any surface, when some of them are brought together, the resulting tube and theories on both sides of it are determined completely. We gave the gauge group along the tube and the pole structures of the created punctures for all the cases (as well as any needed additional information such as the $N' \le N$ in which a resulting theory is defined and the number of additional free hypermultiplets in the corresponding description). \\
We could hope naively that if for instance we bring the punctures on the right side close together,  the pole structures of these punctures alone may be sufficient to fix the additional data required to specify the decoupling result (that is, the gauge group along the tube, the punctures $L$ and $R$, and the $N' \le N$ and additional number of hypers when needed). From what we saw,  it does not have to be the case. An example is shown in \autoref{fig:several_decoupling_options}. There, two full punctures are brought together, but give different tubes.

We may ask when does one side of the decoupling determines the gauge group along the tube and the additional data mentioned above (all of these together will be called shortly "the tube" below). For the sphere, we can get the answer by inspecting the three cases from \autoref{subsection:decoupling_sphere} that cover all the scenarios. Suppose we are given a set of punctures that are brought together, and calculate $\Delta _k$. Begin with the case in which they are of the $SU$ case. If $T$ of that $\Delta _k$ equals $N$, we might be for instance on the left side of the first case, with different possibilities for the right side. In the first case, the tube is determined by the right side as can be seen by the results there, and hence we cannot determine the tube uniquely. If $T=N-1$ we might still be either in the first or the third case, and the tube is not determined. If $T<N-1$ we are necessarily on the right side of the first case, and it determines the tube uniquely. Now suppose that the $\Delta _k$ we obtained is of the $Sp$ type. If $T=N$ even, we might be either in the situation of the second case or of the third one which will give different tubes (in particular different gauge groups). If $T<N$, we are necessarily in the second case, the right side of which fixes the tube. \\
To conclude, a set of punctures $P^i$ brought together determines the information needed to specify the result of the decoupling, when they are of the $SU$ type with $T<N-1$ or the $Sp$ type with $T<N$. As will be explained later on, $T$ can be expressed in terms of the Young diagrams of $P^i$ through \eqref{eq:T_diagrammatic} where $p^1 \ge p^i$ is the puncture with the largest $p$ and $\alpha =\sum _{i \ge 2} p^i$.

For a $g \ge 1$ surface, we have only the first two cases and the punctures that are brought together correspond to the right hand sides there. Therefore the punctures that are brought together determine the tube completely.


\section{Diagrammatic decoupling} \label{sec:diagrammatic_decoupling}

\subsection{Punctures appearing at the end of a tube} \label{section:punctures_at_end_of_tube}

It will turn out instructive to distinguish the class of punctures that can be created when a tube decouples in a weak coupling limit. 

According to what we saw in the discussion of the sphere and $g \ge 1$ surfaces, the three cases mentioned in \autoref{subsection:decoupling_sphere} exhaust all the possibilities.  By examining theses options, the (regular) punctures that appear at the end of a tube are exactly those that are given by the formula
\begin{equation} \label{eq:general_PRP_equation}
l_k = \min(k-1, \sum _{i=1}^m p_k^i)
\end{equation}
where $P^i$ are the (regular) punctures that are brought together on the other side of the tube.

Any puncture that can be expressed as in \eqref{eq:general_PRP_equation} for some set of $m$ punctures, can also be obtained from only two punctures.
A short way to see this, is to note the identity
\begin{equation} \label{eq:several_punctures_relation_to_pairs}
\begin{split}
\min(p_k^1+ \dots p_k^m,k-1)  &= \\
&=\min(k-1,p_k^1 + \min(k-1, p_k^2+ \dots \min(k-1,p_k^{m-1} +p_k^m)\dots )
\end{split}
\end{equation}
for regular punctures $P^1, \dots ,P^m$. Let us just write a particular case of this formula, so that the form of the formula will be clear: \eqref{eq:several_punctures_relation_to_pairs} for $m=3$ is
\begin{equation}
\min(k-1,p_k^1+p_k^2+p_k^3)=\min(k-1,p_k^1+\min(k-1,p_k^2+p_k^3)) .
\end{equation}

Let us prove \eqref{eq:several_punctures_relation_to_pairs}. Denote the term added to each $p_k^i$ by $\bar p_k^i$. Look at $p_k^1+\bar p_k^1$. When $p_k^1+\bar p_k^1 \le k-1$ necessarily we do not choose in the next minimum the $k-1$. So in this region, $p_k^1+\bar p_k^1 = p_k^1+p_k^2+\min(k-1,p_k^3+\bar p_k^3)$. Continuing with the same reasoning, we get that when $p_k^1+\bar p_k^1 \le k-1$, $p_k^1+\bar p_k^1=p_k^1+\dots + p_k^m$. \\
By repeated use of the RHS of \eqref{eq:general_PRP_equation} being regular, $\bar p_k^1, \dots \bar p_k^{m-2} $ are regular punctures. Apply now the analysis of \autoref{section:maximal_gauge_group_along_tube} to $p_k^1$ and $\bar p_k^1$. For $k>T$, $p_k^1+\bar p_k^1 \le k-1$ and we saw what happens for these values of $k$. For $k \le T$, $p_k^1+\bar p_k^1 \ge k-1$. The behavior of $p_k^1+\dots +p_k^m$ is of the $SU$ type only. It implies that $p_k^1+ \dots +p_k^m\ge k-1$. For both of the ranges of $k$ we get 
\begin{equation} \label{eq:several_punctures_relation_to_pairs_proof}
\min(k-1,p_k^1+\bar p_k^1)=\min(k-1,p_k^1+ \dots +p_k^m)
\end{equation}
giving \eqref{eq:several_punctures_relation_to_pairs} indeed.
This shows in particular that the punctures obtained by the RHS of \eqref{eq:several_punctures_relation_to_pairs_proof} can also be achieved using the LHS. 

Call the set of all (regular) punctures $l_k$ that can be written as $l_k=\min(k-1,p_k+p'_k)$ where $p_k$ and $p'_k$ are regular punctures, \textbf{primary regular punctures (PRPs)}. The regular punctures that can appear at the end of a tube are exactly the PRPs. We will give a simple classification of the PRPs now.

\subsection{Classification of PRPs}

Let us give a simple characterization of the possible PRPs in the language of the corresponding Young diagram.
We saw that any PRP can be obtained by colliding two punctures $P$ and $P'$. We defined $T$ to be the last $k$ for which $\Delta _k=p_k+p'_k-k \ge 0$. Let us show that for a PRP $L_1-L_2 \ge T$ and $T \ge L_2$. Assume meanwhile that $T<N$.
After $k=T$ we saw that $\Delta _k$ does not increase anymore, and therefore $k=T+1$ is already at the tip of at least one of the diagrams, say $P'$.
At $k=T+1$ we must have $\Delta _k=-1$, that is $p_k+p'_k=k-1$, and we get $l_{T+1} =T$. $k=T$ is the end of a row in both $P$ and $P'$ (because afterwards $\Delta _k$ decreases by 1). Denote the width of the row coming right after $k=T$ in $P$ by $P_i$. After $k=T$, $p'_k$ will not change anymore (stuck at the tip), so $\Delta _k$ will stay $-1$ as long as we increase $p_k$, that is, $P_i$ more times. We get therefore that $L_1=T+P_i$. Afterwards, $l_k$ becomes $k-2$, and stays so as long as $\Delta _k$ stays the same, which happens for $P_{i+1} $ more steps. Then $L_2=P_{i+1} $. We get that $L_1-L_2=T+P_i-P_{i+1} \ge T$, and $L_2=P_{i+1} \le P_1 \le T$ (the last inequality holds because as long as $k \le P_1$, $p_k=k-1$ and $\Delta _k=p_k+p'_k-k \ge p'_k-1 \ge 0$ so $T \ge P_1$). \\
We assumed that $T<N$. If this does not happen, then $T=N$. In both the $SU$ and the $Sp$ cases, this means that $p_k+p'_k \ge k-1$ for all $2 \le k \le N$. By the PRP formula \eqref{eq:general_PRP_equation}, $l_k=k-1$ for all $2 \le k \le N$. Remembering that $l_k=k-h(k,L)$ ($h(k,L)$ being the row number of the box number $k$ in the Young digram), it implies that $L_1=N$ and $L_i=0$, $i \ge 2$. \\
We obtained that anyway, for a PRP :
\begin{equation} \label{eq:PRP_classification_bounds}
L_1-L_2 \ge T \qquad \text{and} \qquad T \ge L_2
\end{equation}
This implies that $L_1-L_2 \ge L_2$, or $L_1 \ge 2L_2$ is a necessary condition for a PRP.

Now we claim that it is almost sufficient. Given $L_1 \ge 2 L_2$ (except for a simple puncture when $N>2$), it is a PRP : \\
Define $p_k$ through its Young diagram. Take it to have rows with the number of boxes being $L_1-L_2$, $L_2$, $L_2$, $L_3$, $L_4$ and so on. Choose $p'_k$ to be a simple puncture. The values of $p_k+p'_k$ for $k$'s in the corresponding rows of $p_k$ are $k$ in first, $k-1$ in second, $k-2$ in third, and so on (recall $p'_k=1$ for a simple puncture for all $k$, and $p_k$ is $k$ minus the height of the $k$th box in the diagram). $\min(k-1,p_k+p'_k)$ is $k-1$ for the first $L_1-L_2+L_2=L_1$ values of $k$, it is $k-2$ for the next $L_2$ values, $k-3$ for the next $L_3$ values and so on. We therefore constructed $L$ (rows of width $L_1,L_2,L_3, \dots $) using $\min(p_k+p'_k,k-1)$. We required that $L$ is not a simple puncture if $N>2$, because otherwise our construction of $p_k$ gives a no-puncture (all rows have a single box). A simple puncture in $N>2$ is not a PRP, because for instance we saw the requirement $L_1-L_2 \ge T \ge 2$, which is not satisfied by a simple puncture. When $N=2$ a simple puncture is also a full puncture, having $L_2=0$.

To summarize, PRPs are the regular punctures having $L_1 \ge 2L_2$, not including the simple punctures of $N>2$.

\subsection{Diagrammatic construction of the decoupling} \label{subsection:diagrammatic_method}

A (regular) puncture that can appear at the end of a tube, can be obtained by the decoupling of some punctures $P^i$, and will be given by \eqref{eq:general_PRP_equation}. This relation can be described diagrammatically in a simple way. We will show now how the Young diagram of $L$ is found easily from those of the $P^i$.

Recall that a Young diagram of a puncture can end with consecutive rows of width 1, and we called that region of the diagram the tip of the diagram. We also denoted the row number of box number $k$ in the diagram of the puncture $P$ by $h(k,P)$. Note that since $p_k=k-h(k,P)$, $p_k$ equals the number of boxes until box number $k$ that are not in the first column.

\begin{figure}[h]
\centering
\includegraphics[width=0.5\textwidth]{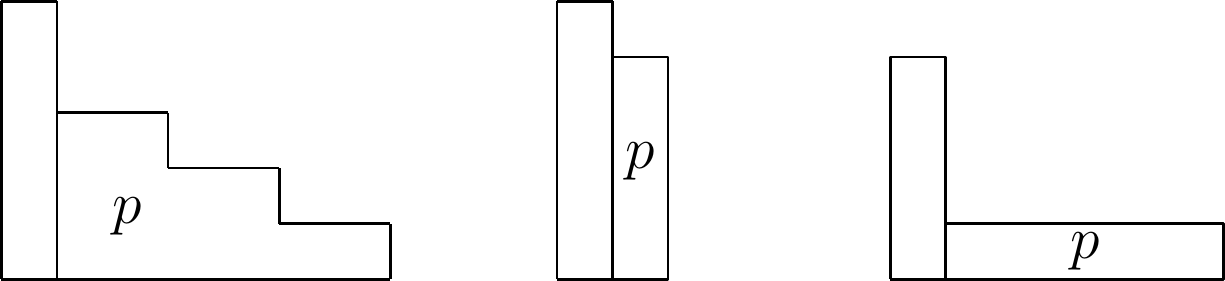}
\caption{Demonstration of the tip of a diagram for a puncture $P$ and the value $p$.}
\label{fig:puncture_p_demo}
\end{figure}

For a general puncture $P$ we have the following restrictions. At $k= \min(2p,N)$ we at least reach the last box of the diagram without the tip, and hence after this $k$, $p_k$ does not change anymore (\autoref{fig:puncture_p_demo} might be useful in this discussion):
\begin{equation} \label{eq:diagrammatic_first_range}
p_k = p \qquad  \text{for } k \ge \min(2p,N) .
\end{equation}
What about the region $k \le \min(2p,N)$? If $P_1=2$ then $p_k = \floor*{\frac{k}{2} }$ for these $k$'s. If $P_1 >2$, at any box in this range at least half of the boxes lay outside the first column, and so $p_k \ge \frac{k}{2} $. Together:
\begin{equation} \label{eq:diagrammatic_second_range}
\begin{split}
p_k = \floor*{ \frac{k}{2}  } \qquad  \text{if } P_1=2, \\
p_k \ge \frac{k}{2}  \qquad \text{if } P_1>2 ,
\end{split} \qquad 
\text{for }2 \le k \le \min(2p,N).
\end{equation}

Arrange the punctures $P^i$ such that $p^1 \ge p^2 \ge p^3 \ge \dots $. Note that for $k \le \min(2p^2,N)$ we have $\sum _i p^i_k \ge k-1$ by \eqref{eq:diagrammatic_second_range}.
Define $\alpha =\sum _{j \ge 2} p^j$.
Suppose that $2p^2 \le N$. Since $2p^2 \le 2p^1$, we have $p^1_{k=2p^2} \ge \floor*{\frac{k}{2} } = p^2$ and the row number of the corresponding box is $h(2p^2,P^1) = 2p^2 - p^1_{2p^2} \le p^2 \le \alpha $. For $k \ge \min(2p^2,N)$, using \eqref{eq:diagrammatic_first_range}, $\sum _{i \ge 2} p^i_k= \alpha $ is constant. In this range, $l_k=\min(k-1,\sum _i p_k^i) = \min(k-1, \alpha + k - h(k,P^1))$. We see that until row number $\alpha +1$ in $P^1$ (which includes the range $k \le \min (2p^2,N)$ indeed since $h(2p^2,P^1) \le \alpha $) $l_k=k-1$, that is we are in the first row of $L$. For $k$ in row number $\alpha + n$ in $P^1$ (with $n > 1$), $l_k=k-n$ so we are in the $n$th row of $L$. In other words, $L$ is just $P^1$ with the $\alpha +1$ first rows combined to a single row. \\
If on the other hand $2p^2 > N$ then at $k=N$ we have $p^1_N \ge \floor*{\frac{N}{2} } =N - \ceil*{ \frac{N}{2} } \ge N - p^2$, or $h(N,P^1) \le p^2$. We saw that for $k \le \min(2p^2,N)$ we have $\sum _i p^i \ge k-1$ and so for all $k$, $l_k=k-1$ which is a full puncture. This is still described by combining the first $\alpha +1$ rows of $P^1$ in the sense that $\alpha +1 \ge p^2 \ge h(N,P^1)$, that is there are more rows to combine than we have in $P^1$.

\begin{figure}[h]
\centering
\includegraphics[width=0.9\textwidth]{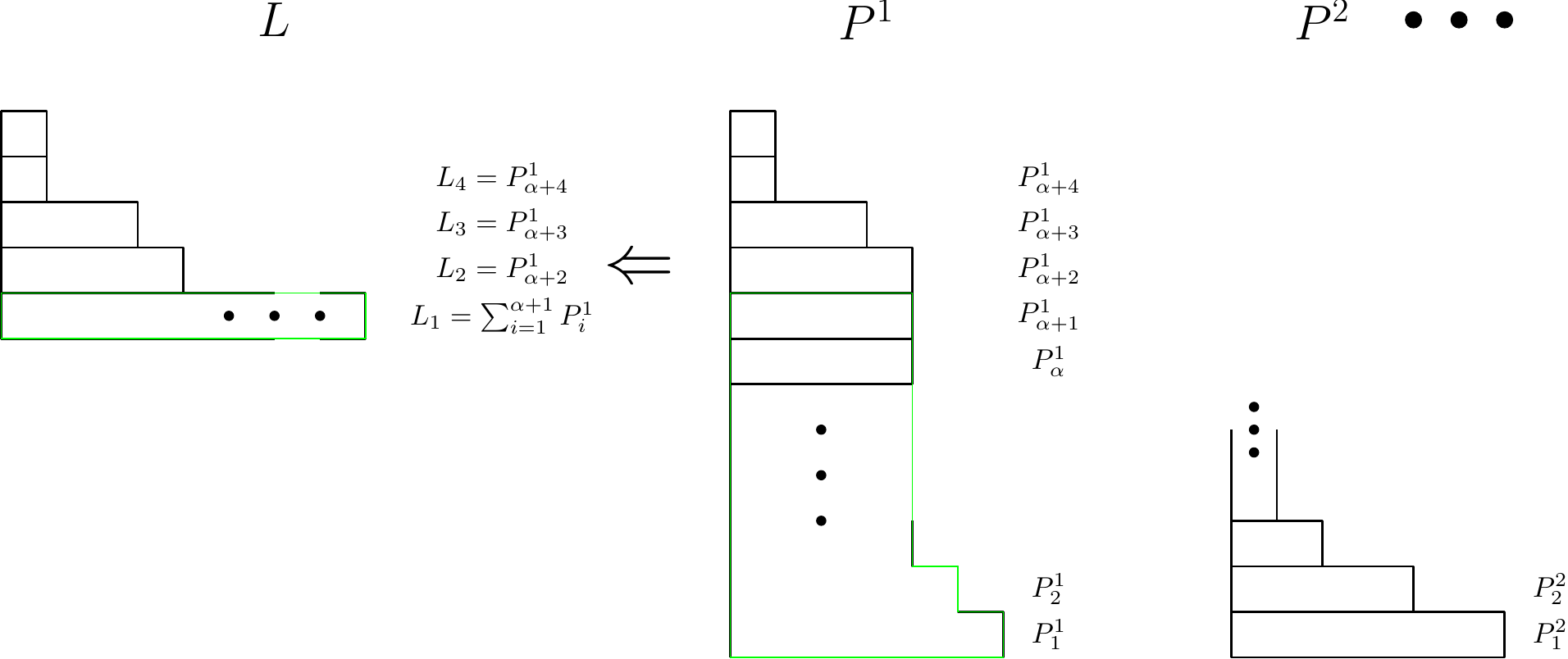}
\caption{The diagrammatic method.}
\label{fig:diagrammatic_decoupling}
\end{figure}

To summarize, arranging the punctures such that $p^1 \ge p^i$ for all $i$ and defining $\alpha =\sum _{i \ge 2} p^i$, the Young diagram of $L$ is obtained by combining the first $\alpha +1$ rows of $P^1$ to a single row. This is demonstrated in \autoref{fig:diagrammatic_decoupling}. If the number of rows in $P^1$ is less than $\alpha +1$, we just combine all of them to a single row, and $L$ is a full puncture.

\subsection{Irregular punctures} \label{subsection:irregular_punctures}

We have given a simple result for the process of decoupling, which requires the use of the familiar regular punctures only. The same applies to the rest of the sections. However, if we would like to describe the resulting theories in the same $A_{N-1} $ theories (that is, the same $N$), we have to use irregular punctures. In this section we pause to discuss them. These irregular punctures are discussed in \cite{Chacaltana:2010ks},\cite{Chacaltana:2011thesis}. A-priori it is not possible to say what is the set of all irregular punctures for $N$ general (for that, it is necessary to check in what 3-punctured spheres they might appear). With the tools we gained, we can now classify all the irregular punctures. We introduce Young diagrams analogous to those of regular punctures. The forms of these diagrams that correspond to irregular punctures will be described.

\begin{figure}[h]
\centering
\includegraphics[width=0.5\textwidth]{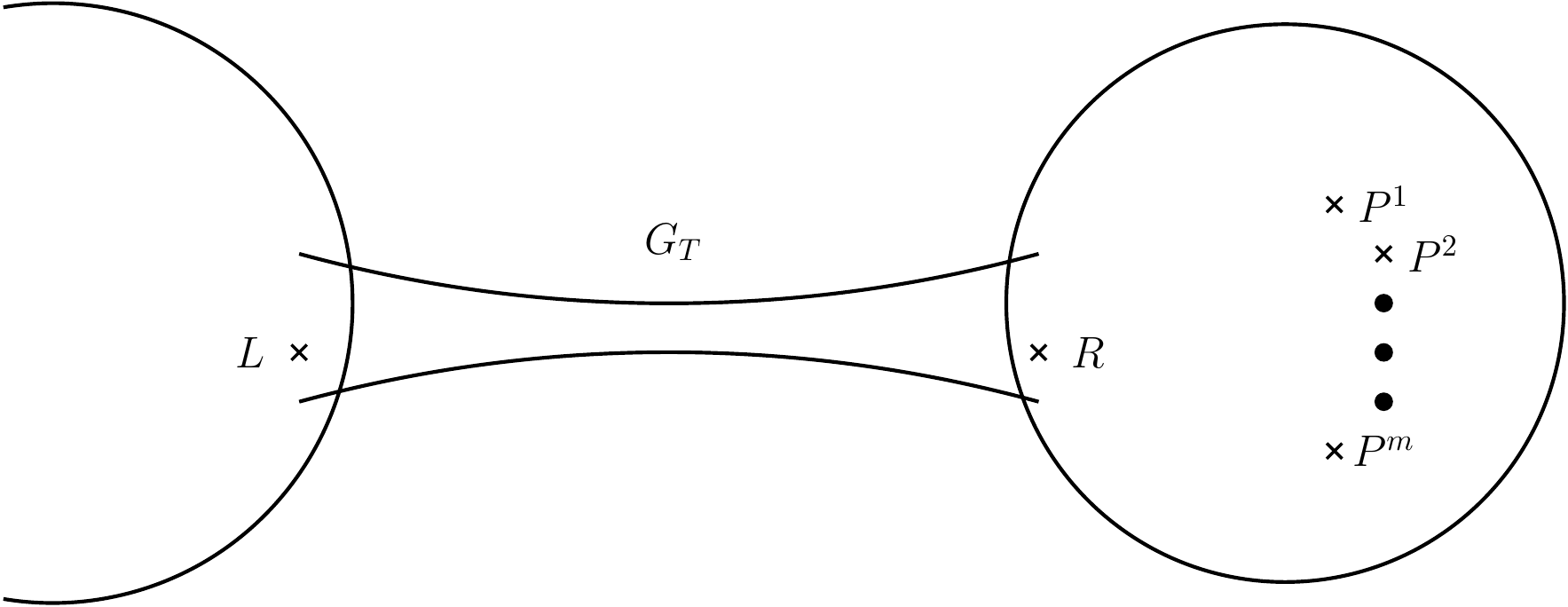}
\caption{Bringing several punctures together.}
\label{fig:bringing_many_punctures_together}
\end{figure}

By considering figures \ref{fig:two_spheres_SUSU},\ref{fig:two_spheres_SUSp1},\ref{fig:two_spheres_SUSp2} we see that all the irregular punctures can be obtained in a decoupling process on the RHS of \autoref{fig:two_spheres_SUSU} or \autoref{fig:two_spheres_SUSp1} (these include the irregular punctures of \autoref{fig:two_spheres_SUSp2}). Therefore, a general irregular puncture can be realized in the setup of \autoref{fig:bringing_many_punctures_together} in which $L$ is a regular puncture (a PRP). Necessarily the puncture $R$ is either an irregular puncture or a full puncture. It will then be convenient to refer to a puncture which is either a full puncture or an irregular puncture, as a \textbf{PIP}. We will study PIPs in this setup of \autoref{fig:bringing_many_punctures_together}.

From figures \ref{fig:two_spheres_SUSU},\ref{fig:two_spheres_SUSp1}, there are two types of irregular punctures, depending on whether the punctures $P^i$ are of the SU type or the Sp type. We will call the corresponding irregular punctures SU and Sp irregular punctures. Their pole structure is given by
\begin{equation}
\begin{split}
SU \text{ irregular puncture : } &p_k=k-1 \text{ for } k \le T<N \text{ and }\\
p_k=2k-1&-\sum_i p_k^i \ge k \text{ for } k>T, \text{ with some set of regular punctures } p_k^i ,
\end{split}
\end{equation}
\begin{equation}
\begin{split}
Sp \text{ irregular puncture : } &p_k \text{ oscillates } k-1,k,k-1,k,\dots,k-1 \text{ up to some even } T \text{ and}\\
p_k=2k-1&-\sum_i p_k^i \ge k \text{ for } k>T, \text{ with some set of regular punctures } p_k^i .
\end{split}
\end{equation}

In \autoref{fig:bringing_many_punctures_together}, while $L$ is given by \eqref{eq:general_PRP_equation}, $R$ can be expressed as (see figures \ref{fig:two_spheres_SUSU},\ref{fig:two_spheres_SUSp1})
\begin{equation} \label{eq:PIP_equation}
\begin{split}
r_k &= \begin{cases}
k-1 & \Delta _k \ge 0 \\
2k-1 - \sum_i p^i_k & \Delta _k < 0
\end{cases}  \\
&=\max(k-1, 2k-1-\sum_i p^i_k) .
\end{split}
\end{equation}
In \autoref{subsection:diagrammatic_method} we gave several arguments, that when combined with \eqref{eq:general_PRP_equation} resulted in the diagrammatic description. We now apply the results there and use \eqref{eq:PIP_equation}. First, consider the case where $P^i$ have the SU behavior. In that case, we can refine the inequality in the range $k \le \min (2p^2,N)$ to $\sum _i p^i_k \ge k$ by \eqref{eq:diagrammatic_second_range} (since either $m >2$ or at least one of the $P^i$ have $P^i_1>2$). For $k \ge \min(2p^2,N)$, $\sum _{i \ge 2} p^i_k = \alpha $. Substituting in \eqref{eq:PIP_equation}, in this range, $r_k = \max(k-1, k-1 - \alpha + h(k,P^1))$.
This equation actually holds for all $k$ since for $k \le \min(2p^2,N)$ we have $k-1-\alpha +h(k,P^1) \le 2k-1 - \sum _i p^i_k \le k-1$.
Therefore, until row number $\alpha $ in $P^1$, $r_k=k-1$, and for a $k$ in row number $\alpha +n$ in $P^1$, $r_k = k-1 + n$.

\begin{figure}[h]
\centering
\includegraphics[width=0.7\textwidth]{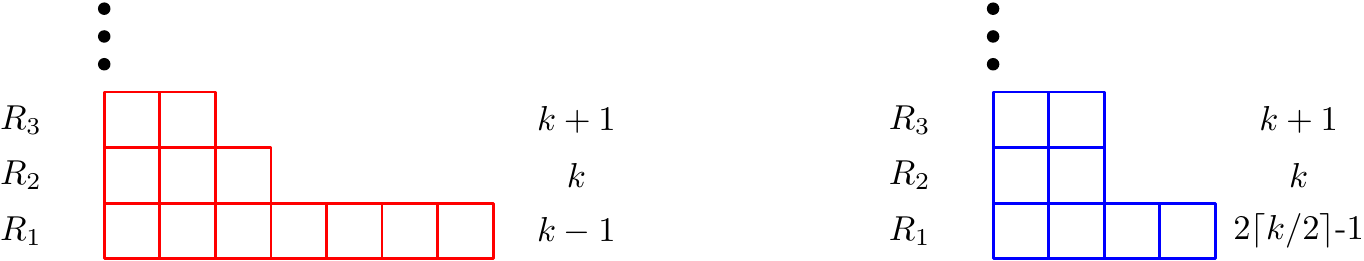}
\caption{Diagrams for PIP (irregular and full) punctures. SU PIPs are marked in red, Sp PIPs are marked in blue.}
\label{fig:irregular_punctures_diagrams}
\end{figure}

We introduce Young diagrams for SU PIPs (that is an SU irregular or a full puncture) as on the left side of \autoref{fig:irregular_punctures_diagrams}. These are usual Young diagrams, colored in red to distinguish them. The box number $k$ in the $n$th row gives $r_k = k-2 + n$. We see by the discussion above that the puncture $R$ of SU PIP type is obtained diagrammatically from the $P^i$'s in a similar way to the regular $L$. Arranging $p^1 \ge p^i$ for all $i$, the diagram of $R$ is that of $P^1$ with the $\alpha $ first rows combined to a single row, $\alpha = \sum _{i \ge 2} p^i$. In a moment we will classify the diagrams that are obtainable.

Before that, let us address the remaining Sp case. We have only $P^1$ and $P^2$, $P^i_1=2$ for both, and $\alpha =p^2$ assuming $p^1 \ge p^2$. Necessarily $2p^2 \le N$ since $P^2_i \le 2$.  In the range $k \le 2p^2$, by \eqref{eq:diagrammatic_second_range}, $\sum _i p^i_k = 2 \floor*{\frac{k}{2} }$. Using \eqref{eq:PIP_equation}, $r_k = \max(k-1, 2 \ceil*{ \frac{k}{2} } - 1) = 2 \ceil*{ \frac{k}{2} } - 1 $. For $k \ge 2p^2$, $p^2_k = p^2$ and $r_k = \max(k-1,k-1-\alpha +h(k,P^1))$. Thus again if $k$ is in row number $\alpha +n$ in $P^1$, $r_k=k-1+n$. Note that $p^1_{k=2p^2} = \floor*{\frac{k}{2} } = p^2$ and $h(2p^2,P^1) = 2p^2-p^1_{2p^2} = p^2 = \alpha $.

We see that for the $\alpha $ first rows of $P^1$, $r_k=2 \ceil*{\frac{k}{2} }-1$, while for the $(\alpha +n)$th row ($n>0$), $r_k=k-1+n$. We represent Sp PIPs (Sp irregular puncture or a full puncture) by a Young diagram, colored in blue, where a $k$ in the first row is associated with $r_k= 2 \ceil*{ \frac{k}{2} } - 1$ and for the $n$th row ($n>1$), $r_k = k-2+n$, see the right side of \autoref{fig:irregular_punctures_diagrams}. Diagrammatically, we get $R$ by just combining the first $\alpha =p^2$ rows of $P^1$ to a single row (where again $p^1\ge p^2$). Since $R_1 = 2\alpha$, the number of boxes in the first row of an Sp PIP diagram is always even.

We are now in a position to classify the PIPs and the irregular punctures. Any PIP is described by a Young diagram from \autoref{fig:irregular_punctures_diagrams}. The question is what subset of these diagrams give precisely all the PIPs. Note that the two types of diagrams have an overlap which is given by the Sp diagrams of $R_1=2$. These diagrams are realized precisely when $m=2$, $P^1_1=2$ and $P^2$ is a simple puncture. To avoid this redundancy, even though this set of $P^i$ is of Sp type, we will group it with the SU PIPs. First, we claim that any SU PIP diagram is obtainable (except for $R_1=1$ as usual). The reason is that we can simply take $P^1$ to be a general regular puncture and $P^2$ to be a simple puncture having $p^2=1$, and then by the diagrammatic description $R$ has the same diagram as $P^1$. The SU PIP diagrams are the same as the regular punctures diagrams (which are all the Young diagrams with more than one column). Next consider Sp PIPs. We have $P^1$ with $p^1$ rows of length 2 and the rest contain a single box, and $P^2$ with $p^2 \le p^1$ rows of 2 boxes and the rest with a single one. By the diagrammatic rules, $R$ has $R_1 = 2p^2$, then $p^1-p^2$ rows of length 2, and the rest with a single box. Thus all the Sp PIPs are such that $R_1$ is even and $R_i \le 2$ for $i>1$. To avoid the redundancy above we restrict to $R_1>2$. 

To recover the irregular punctures we just need to throw away the full punctures, which are the SU PIPs with $R_2=0$. Summarizing, the irregular punctures are described by all the SU Young diagrams with at least 2 rows and columns and by the Sp Young diagrams with $R_1 >2$ even and $R_i \le 2$ for $i>1$. The Sp irregular punctures are described by just two numbers, $p^2 > 1$ and $p^1-p^2 \ge 0$.


\section{Gauging a given theory} \label{sec:gauging}

In the discussion of the decoupling limits we considered, almost always \footnote{Recall the exception for that is the third case mentioned at the end of \autoref{subsection:decoupling_sphere}, which occurs only for even $N$ and gives just $USp(N-2)$. } the Riemann surface in an $A_{N-1} $ theory degenerated to a surface with the same genus and same $N$, with an additional regular puncture $L$ by which it was connected through a tube to a sphere. This situation is demonstrated in \autoref{fig:bringing_many_punctures_together}. We classified what are the regular punctures (the possible $L$'s) that appear at the end of tubes (which we called PRPs). In this section we want to look at this situation from the other direction. Given a PRP $L$, we know it can be gauged, that is, connected through a tube to a sphere with punctures $P^i$. This amounts to gauging a diagonal global symmetry group of the symmetry of $L$ and of the additional sphere. We ask then in \autoref{subsection:gauging_puncture} what are the possibilities for this sphere given a PRP $L$. In \autoref{subsection:possible_gauge_groups} we consider what subgroups of the global symmetry associated with the puncture $L$ can be gauged (that is, what gauge group $G_T$ we can have along a tube). In \autoref{subsection:embedding_in_L} we describe how $G_T$ is embedded in the symmetry of $L$ and what restrictions that were obtained from class-$\mathcal{S} $ are still valid (from field theory considerations) when we replace the additional sphere by a non-class-$\mathcal{S} $ theory. A bound relating the symmetry of $G$ to the symmetries of the punctures $P^i$ is given in \autoref{subsection:bound_sym_PRP}.

\subsection{Gauging a puncture} \label{subsection:gauging_puncture}

Given a PRP $L$, we would like to see how we can find all the possibilities for $P^i$ in \autoref{fig:bringing_many_punctures_together}. For this purpose, the diagrammatic description in \autoref{subsection:diagrammatic_method} (which applies here) will be very useful. We restrict ourselves to $L$ which is not a full puncture, and at the end we address the case of a full puncture. In the discussion in \autoref{subsection:diagrammatic_method} we saw that for $k \ge 2p^2$ (in the ordering of the punctures used there) all $p^i_k$ do not change with $k$ for all $i \ge 2$ (we have used $2p^2 < N$ since $L$ is not a full puncture), and that $k=2p^2$ is in the first $\alpha $ rows of $P^1$. $L$ is obtained by combining the first $\alpha +1$ rows of $P^1$. It follows that after the $\alpha $th row of $P^1$, all other $P^i$ with $i \ge 2$ are at their tip.

Any PRP can be gauged by $2$ punctures as we saw, and therefore we start by addressing the question for $m=2$. Let us denote them instead of $P^1$ and $P^2$, by $A$ and $B$. We order them as in the diagrammatic description with $a \ge b$. We saw that if we restrict their diagrams to the first $L_1$ boxes, we cover exactly the first $b+1$ rows of $A$, and in $B$ we are necessarily at the tip already. Additionally, since $A_{b+2} =L_2$, we must have $A_{b+1} \ge L_2$. The first $L_1$ boxes of $A$ necessarily belong to the following class of diagrams
\begin{itemize}
\item $A(L_1,L_2,n) $ = diagrams of $L_1$ boxes, $n$ rows, and $A_n \ge L_2$
\end{itemize}
with $n=b+1$. The full $A$ diagram is given by removing the first row of $L$ and placing the resulting diagram on top of a diagram in $A(L_1,L_2,b+1)$. The first $L_1$ boxes of $B$ give a diagram in the class
\begin{itemize}
\item $B(L_1,n)$ = diagrams of $L_1$ boxes, $n$ rows, and $B_n=1$ .
\end{itemize}
The full diagram of $B$ is given by extending the tip of this diagram. 

\begin{figure}[h]
\centering
\includegraphics[width=0.9\textwidth]{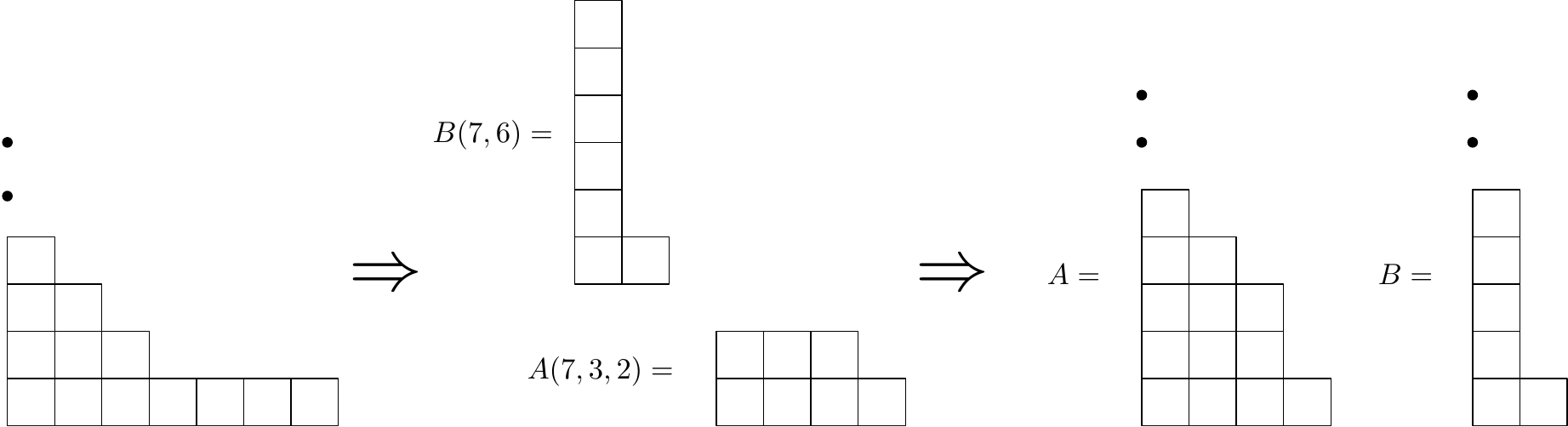}
\caption{Constructing $A$ and $B$. }
\label{fig:puncture_gauging_m2}
\end{figure}

Given any diagram in $A(L_1.L_2,n_A)$ and a diagram in $B(L_1,n_B)$, by extending them as above, we will get in their collision precisely $L$ if, $n_A=b+1$ while $b=L_1-n_B$, that is if
\begin{equation}
n_A + n_B = L_1 + 1.
\end{equation}
For an example, see \autoref{fig:puncture_gauging_m2}.
Note that $n_A \le L_1/L_2$ and so if $L_2 > 1$, then $n_B > n_A$ and $A(L_1,L_2,n_A)>B(L_1,n_B)$ (we sometimes denote $A>B$ meaning $a>b$ for the punctures $A,B$). If $L_2=1$ there will be an overlap between the two classes of punctures. To avoid some of this redundancy we can restrict $n_A \le \frac{L_1+1}{2} $, which ensures that still $a \ge b$.

To encapsulate this part, we do the following to find all the possibilities with $m=2$. For every $2 \le n_A  \le \floor*{ \min \left( \frac{L_1}{L_2} ,\frac{L_1+1}{2} \right) }$ (the upper bound is $\floor*{\frac{L_1}{L_2} }$ unless $L_2=1$), we find all the diagrams in the class $A(L_1,L_2,n_A)$ (there is at least one diagram for each such $n_A$). Each such diagram, completed by appending the rows of $L$ starting from the second one, will be the first puncture. For each of them, we look for all the diagrams of $B(L_1,L_1+1-n_A)$ type, in which we complete the tip to get $N$ boxes in total (there is at least one such diagram for every $n_A$ above). All the pairs of diagrams that were obtained, are the $m=2$ solutions that can appear in a gauging of $L$. If $L_2=1$ we might get the same configuration more than once; for $L_2>1$ we will get each possibility exactly once.

The extension to a general number of punctures $m$ is given in a similar manner. Ordering the punctures as in the diagrammatic description $P^1 \ge P^2 \ge \dots $, the diagrams corresponding to the first $L_1$ boxes are of $A(L_1,L_2,n_1)$ class for $P^1$ and $B(L_1,n_i)$ for $P^i$, $i=2,\dots ,m$. The ordering of the punctures means that $n_1 \le n_2 \le \dots $. To get $L$ in the collision of the $P^i$ we need that $n_1 = \alpha +1 = \sum _{i \ge 2} p^i +1 = \sum _{i \ge 2} (L_1-n_i) + 1$, that is
\begin{equation} \label{eq:gauging_puncture_columns_constraint}
\sum _{i=1} ^m n_i=(m-1)L_1+1.
\end{equation}
For every diagram in $B(L_1,n)$, $n \le L_1-1$ (since a single column diagram is a no-puncture), thus $n_i \le L_1-1$ for $i \ge 2$ and from \eqref{eq:gauging_puncture_columns_constraint} $m \le n_1$. As before, $n_1 \le L_1/L_2 $ for $L_2>1$ and for $L_2=1$, $n_1 \le L_1-1$ again not to have a no-puncture. We have found
\begin{equation} \label{eq:gauging_punctures_allowed_m}
2 \le m \le \frac{L_1}{L_2} -\delta _{L_2,1} .
\end{equation}
Every $m$ in that range is indeed obtainable: take $m-1$ simple punctures, $n_1=m$ with the first row of length $L_1- (m-1)L_2$ and the rest of length $L_2$; all of these punctures are legitimate regular punctures, whose collision gives $L$.

Note that the condition \eqref{eq:gauging_puncture_columns_constraint} ensures that the collision of the punctures $P^i$ gives $L$. If $L_2>1$ then necessarily $n_1 \le n_i$ for $i\ge 2$ (since otherwise $\sum n_i  \le 2 \frac{L_1}{L_2} +(m-2)(L_1-1) \le L_1+(m-2)(L_1-1) = (m-1)L_1 - (m-2) < (m-1) L_1+1 $ which is impossible). Thus indeed $p^1 \ge p^i$ for $i \ge 2$ and the collision gives the combination of the first $L_1$ boxes of $P^1$ giving $L$. If $L_2=1$ then the extension of all the diagrams is with rows of a single box and it does not matter what is the biggest puncture (the combination of the first $L_1$ boxes of all of them gives an L-shaped diagram).

To get all the possible $P^i$ for every $m$ in \eqref{eq:gauging_punctures_allowed_m} we do as before the following. We find all sets of $m-1$ punctures of $B(L_1,n_i)$ class with $n_2 \le n_3 \le \dots \le n_m \le L_1-1$ such that $\sum _{i \ge 2} n_i \ge (m-2)L_1+2$ (these ensure $1 \le m \le n_1 \le L_1-1$). For each of them we look for all $A(L_1,L_2,n_1=(m-1)L_1+1-\sum _{i \ge 2} n_i)$ punctures. Extending the tip of the $B(L_1,n_i)$ punctures to get $N$ boxes in total, and appending to the $A(L_1,L_2,n_1)$ punctures the diagram of $L$ without the first row, we get all the possibilities for the $m$ punctures $P^i$. Note that again if $L_2=1$ we might get the same configurations more than once.

Finally we comment on the case where $L$ is a full puncture. From \autoref{fig:Delta_k_behavior} and \eqref{eq:general_PRP_equation} $L$ is a full puncture exactly when $ \sum _i p^i = \sum _i p^i_N  \ge N-1$. We can get a full puncture with any number $m \ge 2$ of punctures $P^i$. All the possibilities of such punctures with $\sum _i p^i \ge N-1$ result in a full puncture.

\subsection{The possible gauge groups $G_T$} \label{subsection:possible_gauge_groups}

We have found what $P^i$ can appear for a given PRP $L$ in \autoref{fig:bringing_many_punctures_together}. Now we look for all the possible $G_T$.

We have seen in the diagrammatic method that $L$ is given by combining the first $\alpha +1 = \sum _{i \ge 2} p^i + 1$ rows of $P^1$ (where $p^1 \ge p^i$ for all $i$). For the purpose of this subsection, it is more convenient to abuse notation and by $\alpha $ denote $\min\left( \sum _{i \ge 2} p^i ,h(N,P^1) \right)$ (where $h(N,P^1)$ is the number of rows in $P^1$) since there can be less rows in $P^1$ than $\alpha $, in which case $P^1_{\alpha +1} = P^1_{\alpha +2} = \dots = 0$ (note this redefinition is not essential and the original definition of $\alpha $ can be used in the formulae below). The diagrammatic rule is still the same with this definition of $\alpha $. With these conventions, by considering \autoref{section:maximal_gauge_group_along_tube} (in particular \autoref{fig:Delta_k_behavior}) and \eqref{eq:general_PRP_equation} (or alternatively from \autoref{subsection:irregular_punctures} using the last $k$ for which $r_k=k-1$)  one can see that the $T$ associated with the $P^i$ is
\begin{equation} \label{eq:T_diagrammatic}
T= \sum _{i=1} ^{\alpha } P^1_i
\end{equation}
(for both the SU and the Sp case; in the Sp case $p^2 \le h(N,P^1)$ always). We have seen in \autoref{sec:decoupling} that if the $P^i$ are of the SU behavior, then $G_T = SU(T)$ while if they are of the Sp behavior, $G_T = USp(T)$.

From what we have seen just now, $L_1= T+ P^1_{\alpha +1} $, $L_2=P^1_{\alpha +2} $, $\dots $ .Therefore $T=L_1 - P^1_{\alpha +1} $, and we can bound $T$ from both sides. From the usual structure of Young diagrams we have $P^1_{\alpha +1} \ge L_2$ and so $T \le L_1-L_2$. Similarly, $L_1 - P^1_{\alpha +1}  \ge P^1_{\alpha } \ge P^1_{\alpha +1} $, or $P^1_{\alpha +1}  \le L_1/2$ and $T \ge L_1/2$. Additionally, always $T \ge 2$. Combining the bounds
\begin{equation} \label{eq:G_T_rank_bound}
\max\left( 2, \frac{L_1}{2} \right) \le T \le L_1-L_2 .
\end{equation}
Note that as we will see later on, the bound $T \ge \frac{L_1}{2} $ also follows from demanding that the theory on the other side of the tube (not the one with $L$) is unitary.

First we claim that precisely all the $SU(T)$ with $T$ in \eqref{eq:G_T_rank_bound} are the options for $G_T=SU(T)$. Indeed any integer $T$ is obtained in the following configuration. Take $m=2$, $P^2$ a simple puncture, and $P^1$ with rows of length $T$, $L_1-T$, $L_2$, $L_3, \dots$. By \eqref{eq:G_T_rank_bound} both punctures are legal, and they give $G_T=SU(T)$ and $L$ in their collision.

In the Sp case, we have $m=2$ and $P^1_1=P^1_2=2$. Therefore necessarily $L_i \le 2$ for $i \ge 2$. A puncture $P$ with $P_1=2$ is fixed by $p$ --- it has $p$ rows of width $2$ and the rest with width $1$. Suppose $L_2=2$; then necessarily $P^1_i=2$ for all $i \le L_1/2$, $L_1$ is even and $p^2=\frac{L_1}{2} -1 $. In this case $G_T = USp(L_1-2)$. If $L_2=1$, then $P^1_{\alpha +1} $ is $1$ or $2$ and $T=L_1-P^1_{\alpha +1} \ge L_1-2$ and $T \le L_1-1$ from \eqref{eq:G_T_rank_bound}. So if $L_2=1$ and $L_1$ is odd we have $G_T = USp(L_1-1)$ and if $L_1$ is even, then $G_T=USp(L_1-2)$. Both $G_T$ are obtained with a single $P^1,P^2$ configuration. The case $L_2=0$ is completed in the same way, with the result below.

To summarize, for a given PRP $L$, the possible $G_T$ are exactly
\begin{itemize}
\item  $SU(T)$ with $T$ in \eqref{eq:G_T_rank_bound}.
\item If $L_2=2$ and $L_1$ even, $USp(L_1-2)$.
\item If $L_2=1$ then $USp(L_1-2)$ for even $L_1\ge 4$ and $USp(L_1-1)$ for odd $L_1$.
\item If $L_2=0$ and $L_1$ even, can have $USp(L_1)$ and $USp(L_1-2)$ (the latter for $L_1 \ge 4$).
\item If $L_2=0$ and $L_1$ odd, $USp(L_1-1)$.
\end{itemize}
The $Sp$ groups are obtained with a single possibility for $P^1,P^2$, and when $T=2$ the $G_T$ is just $SU(2)$.

\subsection{The embedding of $G_T$ in a PRP} \label{subsection:embedding_in_L}

The bound $T \le L_1-L_2$ in \eqref{eq:G_T_rank_bound} implies that $G_T$ comes from gauging a subgroup of the $U(L_1-L_2)$ factor only in the symmetry \eqref{eq:regular_puncture_symmetry} associated with $L$. As a verification of this, we expect to be left with a $\prod _{i \ge 2} U(L_i-L_{i+1} )$ global symmetry. Indeed, after the gauging, the bigger puncture $P^1$ contains a symmetry of $U(P^1_{\alpha +2} -P^1_{\alpha +3} ) \times U(P^1_{\alpha +3} -P^1_{\alpha +4} ) \times \dots = U(L_2-L_3) \times U(L_3-L_4) \times \dots $ (possibly without a $U(1)$ factor which will appear in another $P^i$) as we saw in the diagrammatic description (see \autoref{fig:diagrammatic_decoupling}). We show in this subsection how the possible $G_T$'s described above are embedded in this $U(L_1-L_2)$ and discuss the possibility of getting additional $G_T$'s if we gauge a diagonal subgroup of the symmetry of $L$ and a non-class-$\mathcal{S} $ theory (replacing the sphere containing the $P^i$). We will see that many of the restrictions on $G_T$ are purely field theoretic, valid for any $\mathcal{N} =2$ SCFT, not necessarily from class-$\mathcal{S} $.

\begin{figure}[h]
\centering
\includegraphics[width=0.5\textwidth]{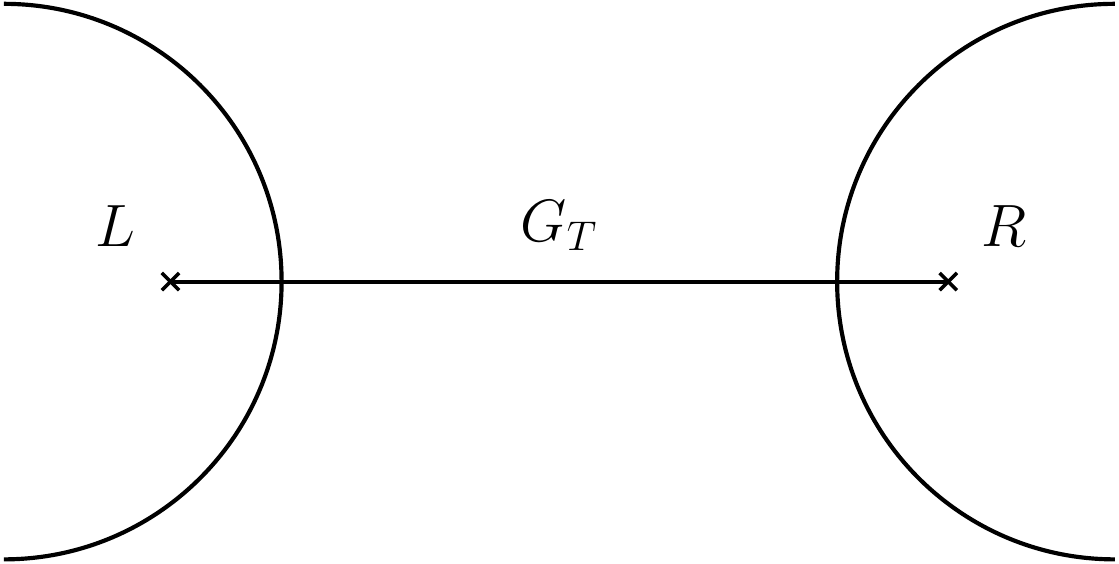}
\caption{The picture of gauging a diagonal flavor symmetry.}
\label{fig:beta_function_picture}
\end{figure}

Any gauge group $G_T$ along some tube in a super-conformal theory should have a vanishing beta-function. Before forming the tube, we had two theories described by two Riemann surfaces (or the tube can have two ends on the same surface). This is depicted in \autoref{fig:beta_function_picture}. This amounts to gauging some diagonal group from the flavor symmetries of the two sides. The contribution of each side to the beta function is proportional to the central charge of the flavor symmetry currents that we gauge. The total contribution of the matter should cancel the contribution from the vector multiplets. The vanishing beta function equation is then
\begin{equation} \label{eq:zero_beta_function}
2 T(\text{adj}) = k^L+k^R
\end{equation}
where $k^L$ and $k^R$ are the contributions from the two sides of the tube. The vector multiplets are in the adjoint representation and therefore contribute $2T(\text{adj})$. The factor of $2$ is a matter of normalization of this equation. We will work in the normalization in which $T(\text{adj})=2K$ and $T(\text{fund})=1$ for $SU(K)$.

\subsubsection{$G_T=SU(T)$}

\begin{figure}[h]
\centering
\includegraphics[width=0.6\textwidth]{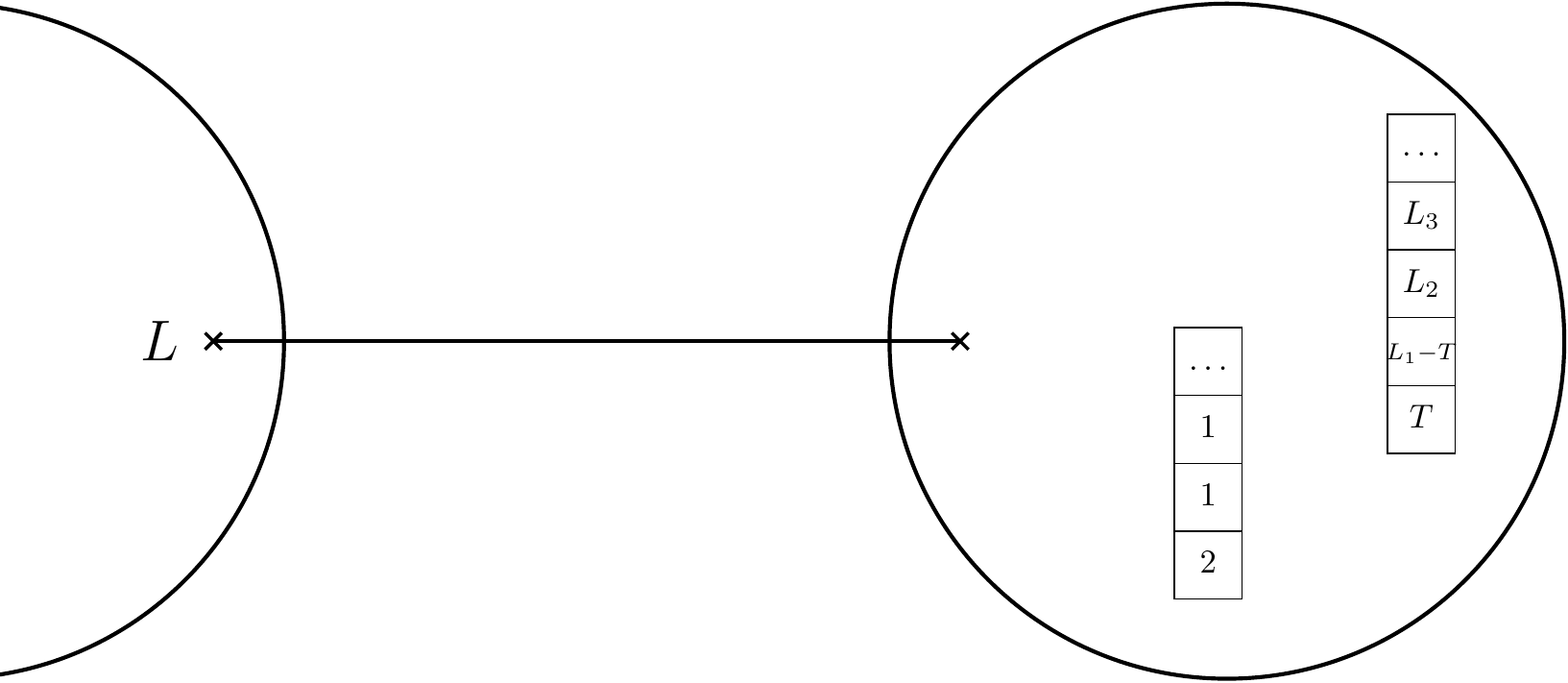}
\caption{Replaced sphere giving $L$ and $G_T=SU(T)$. We used a shorthand notation for a diagram : a diagram with a label $n$ in a box, means that in the appropriate row we have $n$ boxes. The $2,1,1,\dots $ diagram is just a simple puncture.}
\label{fig:embedding_replaced_sphere}
\end{figure}

The central charge of the flavor symmetry of some puncture is determined by the puncture and the subgroup we gauge. We are still in the situation of \autoref{fig:bringing_many_punctures_together}, and suppose $G_T=SU(T)$, $T$ being in the range \eqref{eq:G_T_rank_bound}. 
We can replace the right side of the tube by any other theory, and the contribution of $L$ to the beta function of $G_T$ will not change as long as we have the regular puncture $L$ on the left side of the tube and $G_T$ is the same.
We choose the sphere shown in \autoref{fig:embedding_replaced_sphere} for that. Note that the rightmost diagram is valid, $T \ge L_1 - T$ and $L_1-T \ge L_2$. Using the analysis in \autoref{subsection:decoupling_sphere} or the diagrammatic method, it is seen immediately that we indeed get $L$ and $G_T=SU(T)$.

\begin{figure}[h]
\centering
\includegraphics[width=0.7\textwidth]{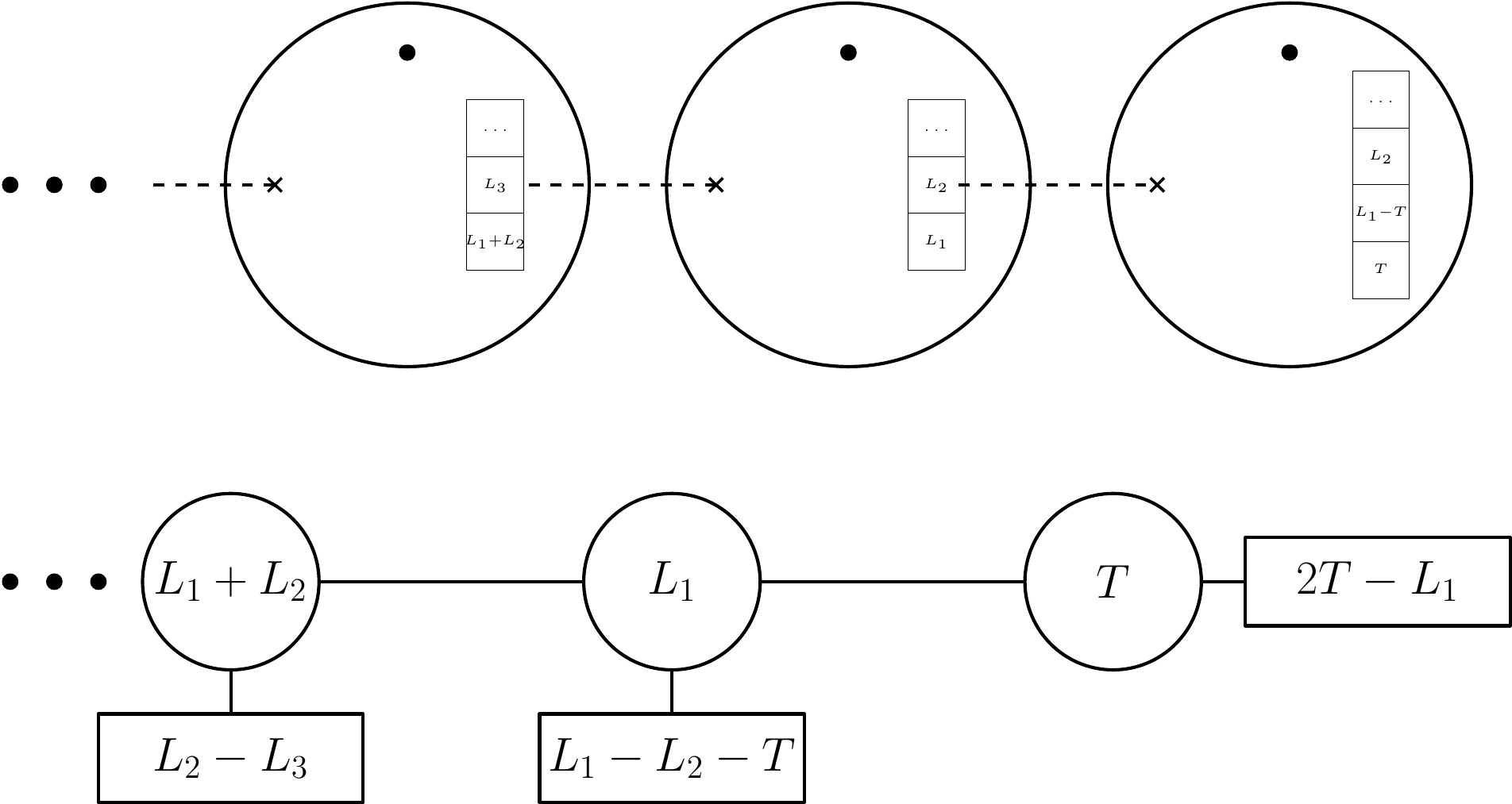}
\caption{A linear quiver in which the same $L$ and $G_T$ appear. A small disk represents a simple puncture.}
\label{fig:embedding_replacement_in_quiver}
\end{figure}

We can get the contribution of $L$ since the configuration of \autoref{fig:embedding_replaced_sphere} appears in the linear quiver shown in \autoref{fig:embedding_replacement_in_quiver}. The rightmost sphere is just a free theory of $2T-L_1$ fundamentals of $SU(T)$, while the contribution of $L$ to $SU(T)$ is the same as of $L_1$ fundamentals (the sphere to which it belongs gives a bifundamental hypermultiplet).
If we have hypermultiplets in the representation $ \oplus _i r_i$ of some group $G$, then $k_G=\sum _i 2T(r_i)$.
Therefore
\begin{equation}
k^L_{SU(T)} =2L_1 T(\text{fund})=2L_1 .
\end{equation}

We saw that any gauging of $L$ must come from some subgroup of $U(L_1-L_2)$. In general, if the central charge of a current of flavor symmetry $U$ ($U$ being a simple group) is $k_U$, then for a simple subgroup $W \subset U$, $k_W =xk_U$ where $x$ is the embedding index of $W$ in $U$. In our case, we found that $k_{SU(T)} ^L = k^L_{SU(L_1-L_2)}$. This means that the embedding index is $1$, and the embedding is the trivial embedding.

We can again check if this is consistent. After gauging $SU(T)$ from the $SU(L_1-L_2)$ part of the symmetry of the puncture $L$, we expect that if the embedding is trivial, there will be a leftover $SU(L_1-L_2-T)$ symmetry. This is indeed the case, as the bigger puncture in the diagrammatic picture of gauging, has an $SU(P^1_{\alpha +1} -P^1_{\alpha +2} )= SU(L_1-T-L_2)$ factor in its symmetry (as we saw in the previous subsection, with $\alpha $ defined as there), see \autoref{fig:diagrammatic_decoupling}.

\begin{figure}[h]
\centering
\includegraphics[width=0.6\textwidth]{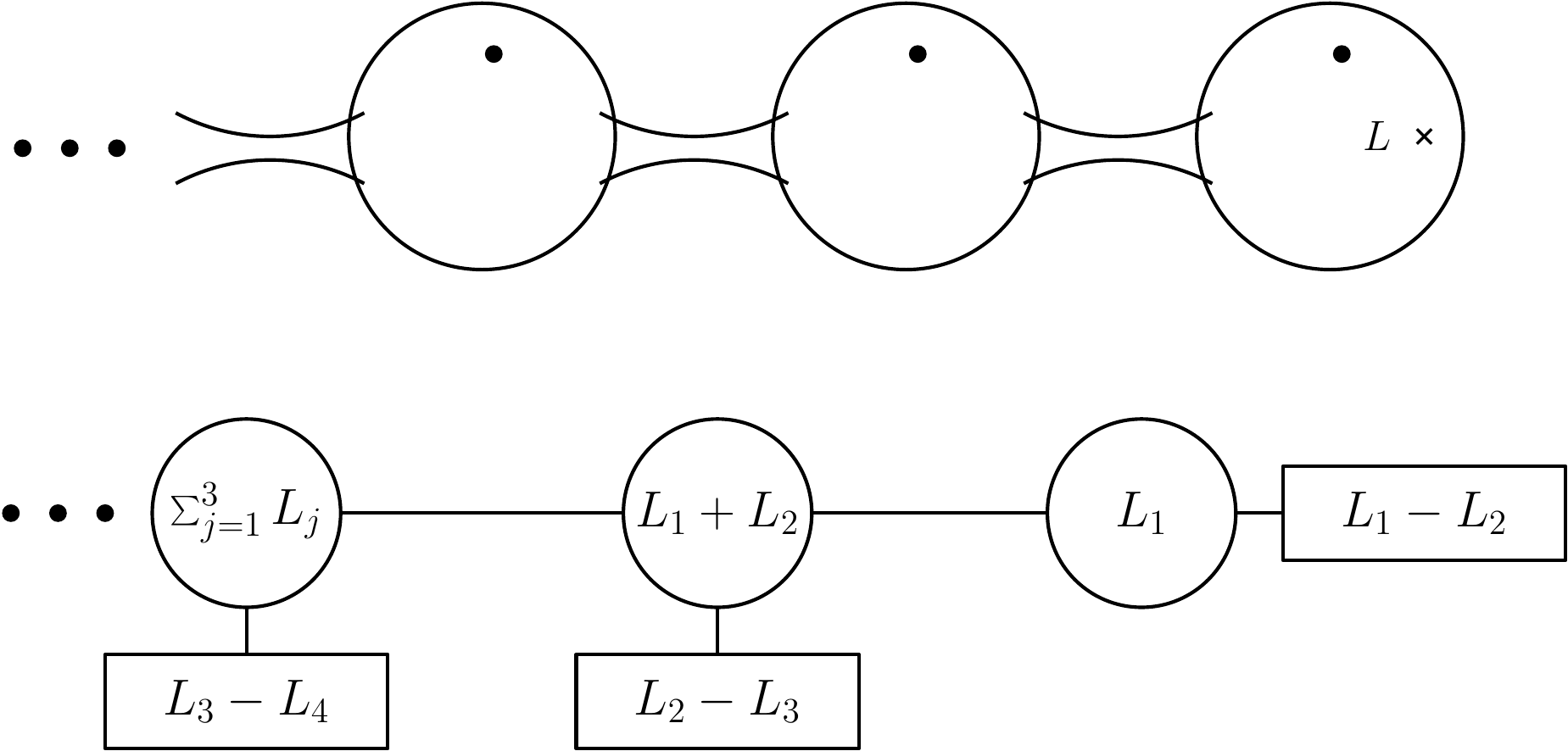}
\caption{A tail of a general linear quiver. In the top we see the Riemann surface with a puncture $L$ and additional simple punctures. Below the corresponding linear quiver is displayed.}
\label{fig:general_quiver}
\end{figure}

We could ask whether in gauging $L$ with a non class-$\mathcal{S} $ theory, we could obtain a gauging which is not possible in class-$\mathcal{S} $, that is some other gauged subgroup or a different embedding (a non trivial embedding). Suppose we could gauge an $SU(T)$ subgroup of $SU(L_1-L_2)$ with embedding index $x$, and at the other end of the tube we could have any other theory (not only of class-$\mathcal{S} $). Then the contribution of the RHS theory to the beta function is
\begin{equation}
k^R = 4T-k^L = 4T - 2L_1 x .
\end{equation}
We assume that the theory at the other end of the tube is unitary. The condition $k^R \ge 0$ means $T \ge L_1/2$. Therefore, we cannot gauge in general an $SU(T)$ that does not appear in class-$\mathcal{S} $. \\
Since the Dynkin embedding index is an integer, for a non-trivial embedding $x \ge 2$, implying $k^R \le 4(T-L_1)$. We saw that $T=L_1-P_{\alpha +1}$ and so $T-L_1<0$ unless $P_{\alpha +1}=0$, for which  $T=N=L_1$. But, for $T=N=L_1$, $x=1$. We see that assuming $x \neq 1$ leads to $T-L_1<0$ and $k^R<0$.
We conclude that the only possible gauging of $SU(T) \subset SU(L_1-L_2)$ is by the trivial embedding. \\
Note that from the same reason, we cannot gauge any subgroup of the other $SU(L_i-L_{i+1} )$, $i>1$, symmetries as in class-$\mathcal{S} $. 
For any $L$ we can form the linear quiver presented at the bottom of \autoref{fig:general_quiver}. The corresponding curve is the one at the top of \autoref{fig:general_quiver}.
If we turn off the gauge coupling of the $SU(\sum _{j=1} ^i L_j)$ gauge group, we find that $k_{SU(L_i-L_{i+1} )} = 2 \sum _{j=1} ^i L_j$. By the supersymmetry, this central charge is also an anomaly coefficient and is independent of the exactly marginal couplings. $k^R \ge 0$ together with $T \le L_i-L_{i+1} $ are impossible as can be seen easily.

\subsubsection{$G_T=USp(T)$}

The only other possibility for $G_T$ in class-$\mathcal{S}$ is $USp(T)$. We saw in the previous subsection the cases giving $USp(T)$.
Let us see what are the possible embeddings given that we cannot have $k^R<0$. For $USp(2r)$ , $2T(\text{adj})=4(r+1)$. The options giving $USp(2r)$ are
\begin{itemize}
\item $L_1=2r+2$, $L_2=0,1,2$. In these cases, $k^R=4(r+1)-x k_{SU(L_1-L_2)} =4(r+1)-2(2r+2)x$. Only $x=1$ is possible, and it gives $k^R=0$.
\item $L_1=2r+1$, $L_2=0,1$. $k^R=4(r+1)-2x(2r+1)$. For $x=1$ we get $k^R=2$. If $x \ge 2$, $k^R \le -4r < 0$, so it is not possible.
\item $L_1=2r$, $L_2=0$. $k^R=4(r+1)-4rx $. For $x=1$ we get $k^R=4$. If $x \ge 2$, $k^R \le 4(1-r) <0$, which is not possible again.
\end{itemize}

For the $USp(2r)$ case, again only the trivial embedding in $SU(L_1-L_2)$ is possible. We already saw that there will indeed be a leftover $SU(L_1-L_2-T)$ symmetry after the gauging.

Let us ask once again about gauging $L$ with a non class-$\mathcal{S}$ theory.
As before, only an embedding with index $x=1$ is possible, even when the other theory that we gauge is not necessarily from class-$\mathcal{S} $. The reason for this is the following. An embedding of $USp(2r) \subset SU(L_1-L_2)$ implies $L_1 \ge 2r+L_2$. Again we use $k^R = 4(r+1)-2xL_1 \le 4(r+1)-2x(2r+L_2)$. An embedding with $x \ge 2$ would imply $k^R<0$ (using $r>1$ since otherwise the algebra is the same as that of $SU(2)$). We saw that in class-$\mathcal{S} $ we cannot gauge some $USp(2r)$ of the flavor symmetry of a puncture with $L_2 \ge 3$, and this still holds even when the other theory we couple to the gauge group is not from class-$\mathcal{S} $ by the same bound.

\subsubsection{$G_T=SO(T)$}

In the theories discussed here, it was not possible to have an $SO(n)$ group along the tube. Let us see that it is still impossible even if we would like to gauge a diagonal subgroup of the symmetry of the puncture $L$, and of some other theory which is not necessarily of class-$\mathcal{S} $. $2T(\text{adj}) = 4(n-2)$ for $SO(n)$. Additionally, in any embedding $SO(n) \subset SU(L_1-L_2)$, $x \ge 2$ necessarily (since the minimal index of a representation is 2 in $SO(n)$). 
The exceptions for that are the low rank cases in which the two algebras are just the same, but we are not interested clearly in these cases.
In all the other cases, $x=2$ and $L_1-L_2 \ge n$. We obtain then $k^R = 4(n-2) -2x L_1 \le 4(n-2) - 4n <0 $.

\subsection{A bound on the rank of the symmetry of a PRP} \label{subsection:bound_sym_PRP}

Given $m\ge 2$ regular punctures $P^1, \dots ,P^m$, the regular puncture $L$ defined by \eqref{eq:general_PRP_equation} satisfies
\begin{enumerate}
\item $L_1 \ge \min( \sum _{i=1} ^m P_1^i - m+1 , N)$ or equivalently
\begin{equation} \label{eq:bound_rk_l_1}
 \rk G (L) \ge \min(  \sum _{i=1} ^m \rk G(P^i), N-1)
\end{equation}
(where $G(P)$ is the global symmetry of the puncture $P$).
\item If in addition at least one $P_2^i \ge 2$, then\\
$L_1 \ge \min( \sum _{i=1} ^m P_1^i - m +2 , N)$ or equivalently
\begin{equation} \label{eq:bound_rk_l_2}
 \rk G(L) \ge \min( \sum _{i=1} ^m \rk G(P^i) + 1 , N-1)
\end{equation}
\end{enumerate}

These follow easily from equation \eqref{eq:general_PRP_equation}.
In any puncture $P_1 \ge 2$, because if it was $1$ it was a no-puncture. Define $k_0 = \min(\sum _i P^i_1 - m+1,N)$. If $k_0 = \sum _i P^i_1-m+1$, then for any $i$, $k_0 = P^i_1+\sum _{j \neq i} P^j_1-m+1 \ge P^i_1 + 2(m-1)-m+1 \ge P^i_1$ and therefore $p^i_{k_0} \ge P^i_1-1$. This is clearly true also if $k_0=N$. It follows that $\sum _i p^i_{k_0}  \ge \sum _i P^i_1 - m \ge k_0 -1$, and then by the definition of $l_k$, $l_{k_0} = k_0-1$. This means that $L_1 \ge k_0$, establishing \eqref{eq:bound_rk_l_1} (since $\rk G(P)=P_1-1$).

For the second statement, define $k'_0 = \min( \sum _i P^i_1 - m+2,N)$. It is not smaller than the previously defined $k_0$ and therefore still $p^i_{k'_0} \ge P^i_1-1$ for every $i$. For the particular $i$ for which $P^i_2 \ge 2$, if $k'_0=\sum _j P^j_1-m+2$ then $k'_0 =  P^i_1 + \sum _{j \neq i} P^j_1 - m + 2 \ge P^i_1 + 2(m-1)-m+2 = P^i_1 +m \ge P^i_1 + 2$. Since $P^i_2 \ge 2$, it means that $p^i_{k'_0} \ge P^i_1$. If $k'_0=N$ this clearly still holds. Together, these give $\sum_i p^i_{k'_0} \ge \sum P^i_1-m+1 \ge k'_0 - 1$, and once again $L_1 \ge k'_0$.


\section{Large $N$ and field theories on $AdS_5$} \label{sec:large_N}

Certain theories of the sort we have used are relevant in the large $N$ limit for describing field theories on $AdS_5$ with various boundary conditions. In this section we address this connection and apply the tools from the previous sections.

In \cite{Aharony:2015zea} the AdS/CFT correspondence between type IIB string theory on $AdS_5 \times S^5 / \mathbb{Z}_K $ and the four dimensional $\mathcal{N} =2$ theory with a circular quiver $SU(N)^K$ \cite{Kachru:1998ys} was considered. The four dimensional side is the theory where the Riemann surface $C$ is a torus with $K$ simple punctures. The singular limit where the integrals of the NS-NS and R-R 2-forms on the 2-cycles of the orbifold vanish is particularly interesting. This corresponds in the 4d side to bringing the simple punctures close to each other. In the large $N$ limit, we get the six-dimensional $(2,0)$ $A_{K-1} $ theory on $AdS_5 \times S^1$ decoupled from gravity.

In the 4d picture, this situation is a simple instance of the degeneration limits we considered in the previous sections. As the simple punctures are taken close, a sphere bubbles off, connected by a tube to the remaining torus on which a single puncture is created if the tube is completely decoupled. This puncture is an L-shaped puncture with rows of width $K+1,1,1,\dots $ and pole structure $1,2,\dots ,K,K,\dots $. The gauge group that becomes weakly coupled is an $SU(K)$. As we explained in \autoref{sec:decoupling}, the curve on the decoupling sphere terminates within a number of terms of the order of $K$. The sphere and the tube are independent of $N$.

In this limit where we have an $A_{K-1} $ singularity in the type IIB theory, we expect to get an $SU(K)$ gauge symmetry in the five dimensional theory on $AdS_5$ and an $SU(K)$ global symmetry in the corresponding four dimensional theory. This does not happen and the global symmetry in our 4d theory is $U(1)^K$. This conflict is avoided as described in \cite{Aharony:2015zea} by having the $SU(K)$ tube on the boundary gauging both the decoupling sphere with simple punctures sitting on the boundary as well, and the five dimensional theory on $AdS_5$ in the bulk. The 4d $SU(K)$ symmetry is thus gauged and not global, and the global $U(1)^K$ symmetry arises from the global symmetry of the theory on the decoupling sphere.
Having these theories on the boundary provides a specific choice of boundary conditions for the $(2,0)$ theory on $AdS_5\times S^1$, which is realized in the type IIB construction. We could give the $(2,0)$ theory on $AdS_5 \times S^1$ alternative boundary conditions in which we do not have the ingredients just mentioned on the boundary. In this case the corresponding four dimensional theory is a torus with the L-shaped puncture, having now $SU(K)$ in its global symmetry. It is not known how to realize these alternative boundary conditions in type IIB string theory. In principle an M-theory dual can be found along the lines of \cite{Gaiotto:2009gz}, but it is not reliable due to high curvatures.

\begin{figure}[h]
\centering
\includegraphics[width=0.6\textwidth]{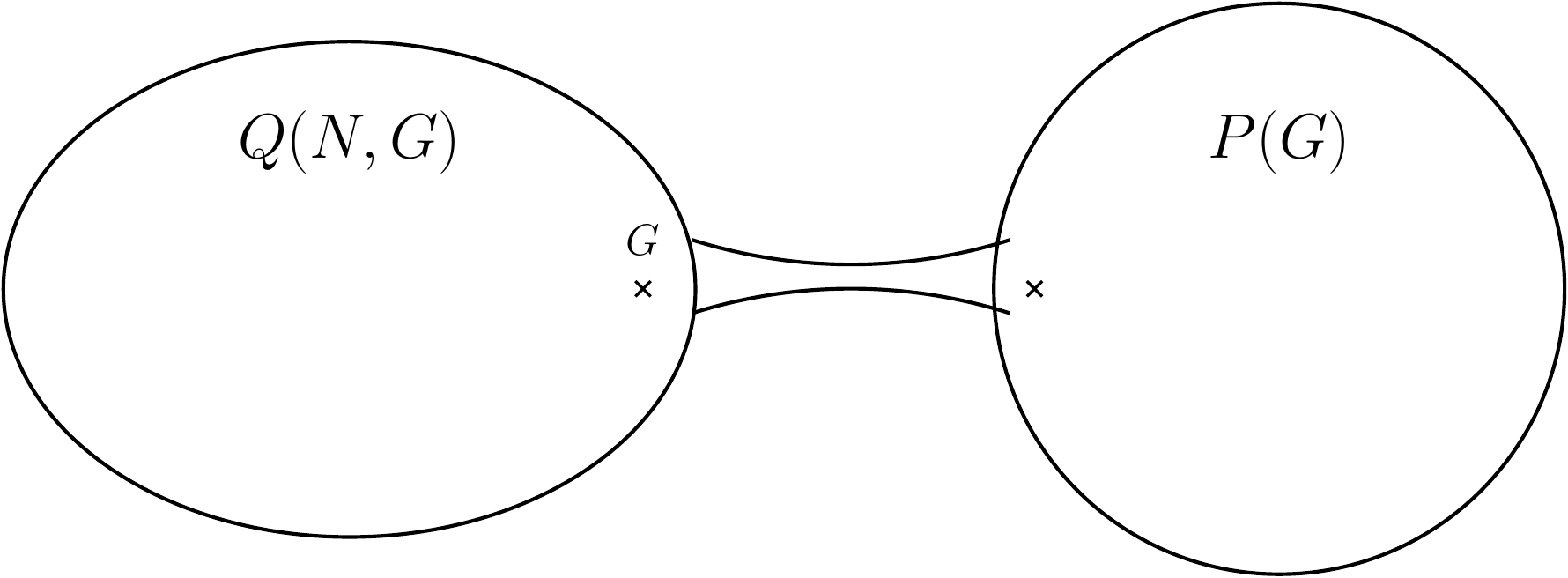}
\caption{General form of 4d theories which might give decoupled theories on $AdS_5$.}
\label{fig:4d_part_correspondence}
\end{figure}

We would like to see by how much this picture can be generalized. The configurations we need to examine appear in \autoref{fig:4d_part_correspondence}. The $Q(N,G)$ ingredient is the analog of the torus with the L-shaped puncture. It is now some surface including a puncture with global symmetry $G$ (as well as additional punctures possibly). The tube, gauging some subgroup of $G$, together with the $P(G)$ theory may provide different boundary conditions in case we have a decoupled theory on $AdS_5$. The tube and the $P(G)$ theories are finite in the large $N$ limit.

In order for the tube and $P(G)$ to be independent of $N$, we must be either in the first or the second general cases which were described at the end of \autoref{subsection:decoupling_sphere}. In particular as was discussed in \autoref{subsection:decoupling_g_ge1}, $P(G)$ must be defined on a sphere. We are once again in the situation we inspected a lot, which appears in \autoref{fig:bringing_many_punctures_together} with the puncture of symmetry $G$ being $L$.

As was mentioned above, a torus with an L-shaped puncture contains a decoupled $(2,0)$ theory on $AdS_5\times S^1$ \cite{Aharony:2015zea}. We first try to see in what theories $Q(N,G)$ we have a decoupled field theory on $AdS_5$. We will give a partial answer for this, not covering all the possible situations. For the cases where this does happen, we assume that different choices of the gauged subgroup $G_T \subset G$ and of $P(G)$ provide different boundary conditions for the decoupled theory. Under this assumption we can enumerate what are those possible boundary conditions.

\subsubsection*{Large $N$}

The first immediate generalization of an L-shaped puncture is a puncture with a tip (all rows are of width $1$ starting from some row number).
When $N$ is large, we will say that a puncture has a long tip if $p$ is independent of $N$.
All punctures of this sort have a finite large $N$ limit for their global symmetry. That is, only for these punctures the global symmetry \eqref{eq:regular_puncture_symmetry} does not change as we append a box (or any number of boxes) starting from some point, and is independent of $N$. As will be explained in a moment, if all the punctures $P^i$ on $P(G)$ before the decoupling of the tube have a long tip in the large $N$ limit, after the decoupling the $P(G)$ side will represent a theory which is independent of $N$.

By the requirement that after decoupling $P(G)$ is independent of $N$ we are led to consider additional punctures. Namely, punctures of "finite $P_1$" which means that the first row of the Young diagram has a number of boxes independent of $N$. We will now explain that except for exceptions of a certain type, the theory $P(G)$ after its decoupling will be independent of $N$ when all punctures except one have a long tip, and the remaining puncture is of "finite $P_1$". Therefore the punctures $P^i$ we need to consider are of "finite $P_1$" (which includes punctures with a long tip in particular).

The explanation of the statement above is the following. As mentioned before, the $P(G)$ theory corresponds to the RHS in the first or the second general cases which were described at the end of \autoref{subsection:decoupling_sphere}. We will use the (by now) usual convention of $p^1 \ge p^i$ for $i>1$ and $\alpha =\sum _{j \ge 2} p^j$. By the description of the second case in which we have only $m=2$, we basically need $\alpha $ to be independent of $N$ for $P(G)$ to be independent of $N$ and so $P^2$ has a long tip and $P^1$ automatically has "finite $P^1_1$" (it equals to $2$). There is an exception for that, in which we have $p^2=\alpha $ which is large and $P^1_{\alpha +1} =2$. In this case the decoupled theory is an empty theory.
Next, consider the first general case. Assume first that there is a $\Delta _k>0$ (the $\Delta _k$ here corresponds to the $P^i$ punctures). To remain with a finite theory we must have a finite $N'$ (see the discussion regarding $N'$ in \autoref{subsection:decoupling_sphere}) which means that $P^1_1$ and $\alpha $ are necessarily independent of $N$ and so $P^i$ for $i>1$ have a long tip. The number of additional free hypers in the resulting decoupled theory will also be finite. Finally consider what happens if there is no $\Delta _k>0$. As mentioned in \autoref{subsection:decoupling_sphere}, the decoupling theory is a theory of free hypers. 
For the number of hypers (given by \eqref{eq:free_hypers_SU_RHS_when_delta_k_le_0}) to be independent of $N$ we need in principle that again $P^1_1$ and $\alpha $ are finite and so again $P^1$ is of "finite $P^1_1$" and the $P^j$ for $j>1$ have long tips. The exception for this is when $m=2$, $P^2$ is a simple puncture, and $P^1_1=P^1_2$ which is large. In this case the decoupling theory is empty as in the previous exception. To summarize, we have found that $P(G)$ is independent of $N$ after decoupling exactly when $P^1$ is of "finite $P^1_1$" and the other $P^i$ have long tips, with the exceptions mentioned above in which the decoupled theory is empty.

Note that we cannot have $Q(N,G)$ a sphere with only a single puncture of symmetry $G$ in \autoref{fig:4d_part_correspondence}, because the number of Coulomb branch parameters of dimension $k$ in the sphere composed of $Q(N,G)$, $P(G)$ and the tube is $d_k=\Delta _k-k+1$, and for $P^i$ giving $P(G)$ finite, $\Delta _k$ and $d_k$ become negative. On a torus, and higher genus surfaces, the Coulomb branch graded dimension \eqref{eq:Coulomb_branch_graded_dimension} is always non-negative, $d_k \ge 0$, and we do not have this restriction.

\subsection{Decoupled field theories in the large $N$ limit} \label{subsection:decoupled_AdS5}

Consider a genus $g>1$ surface $C$ having punctures with a long tip. We would like to see if analogously to the case of a torus with an L-shaped puncture, there is in the large $N$ limit a field theory which is decoupled from gravity. Gaiotto and Maldacena constructed the gravity duals of such theories \cite{Gaiotto:2009gz}. These are given by M-theory on a background of the form $AdS_5 \times \mathcal{M} _6$ with $\mathcal{M} _6$ being compact. The M-theory spacetime is smooth for $g>1$, and in the large-$N$ limit adding to it punctures of the sort we consider is a small local deformation of the background.

The background is found by solving a Toda equation with boundary conditions specified by the different punctures.
Around each puncture, the Toda equation is analogous to three dimensional electrostatics with a cylindrical axial symmetry. Every puncture is associated with a line charge density profile denoted by $\lambda (\eta )$ which is piecewise linear. The slopes are integer and change at integer values of $\eta $. Specifically, for a given puncture $P$, the slopes are $P_1,P_2,\dots $ and they change at $\eta =1,2,\dots $. Whenever $P_i-P_{i+1} \ge 2$, we have corresponding to a slope change, in the appropriate position in spacetime, an $A_{k-1} $ singularity, with $k=P_i-P_{i+1} $. Each $A_{k-1} $ singularity gives a non-abelian gauge theory on $AdS_5$ which is associated with an appropriate global symmetry in the four dimensional $\mathcal{N} =2$ theory. The radius of the $AdS_5$ in Planck units is $R_{\text{AdS}_5} /l_P \sim \lambda (\eta _k)^{1/3} $ (see equation (3.22) in \cite{Gaiotto:2009gz}) where $\lambda (\eta _k)$ is the value of $\lambda $ where the slope changes, and the five dimensional gauge coupling is $R_{AdS_5} /g_5^2\sim \lambda (\eta _k)$.  To get a non-trivial theory decoupled from gravity we need to have a finite non-zero value of $R_{AdS_5} /g_5^2$ while $l_P \to 0$ (compared to $R_{AdS_5} $ and $g_5^2$).
Therefore we see that we do not have a non-trivial field theory decoupled from gravity.

We do not provide indication for or against the existence of decoupled field theories on $AdS_5$ in the large $N$ limit in the rest of the cases. In $g>1$ surfaces having only punctures of the type we considered there are no such decoupled theories, but if there are other punctures, different arguments should be used \footnote{For any theory having an M-theory dual which is given by a weakly curved spacetime with $A_{k-1}$ singularities, there are no such decoupled field theories on $AdS_5$, as was argued. }. In a $g=1$ surface with an L-shaped puncture there is a decoupled theory, but for other punctures, as well as for the sphere, the situation should be investigated further. 

Let us assume that in the cases where we do have a decoupled field theory, the various choices for $P(G)$ in \autoref{fig:4d_part_correspondence} with the appropriate tube amount to different boundary conditions for the decoupled theory on $AdS_5$.
As was mentioned before, $P(G)$ is defined on the sphere and we are in the situation of \autoref{fig:bringing_many_punctures_together} with $L$ having symmetry $G$. We can use all the conclusions developed so far.

Under this assumption, \autoref{subsection:gauging_puncture} essentially is a classification of all the possible boundary conditions of this sort.  In particular, this can be applied to the torus with a single L-shaped puncture where we know that we have a decoupled field theory.

If we do not restrict to boundary conditions which can be implemented in class-$\mathcal{S} $, the possible options for gauging some subgroup of $G$ depend only on the central charge of this subgroup,
which is related to the central charge of $G$ as review above. The options for possible boundary conditions are thus determined by the central charge of $G$, which is related to the gauge coupling of $G$ in the  bulk in units of the $AdS$ radius.

\subsection{Non-singular weakly curved gravitational duals}

\begin{figure}[h]
\centering
\includegraphics[width=0.35\textwidth]{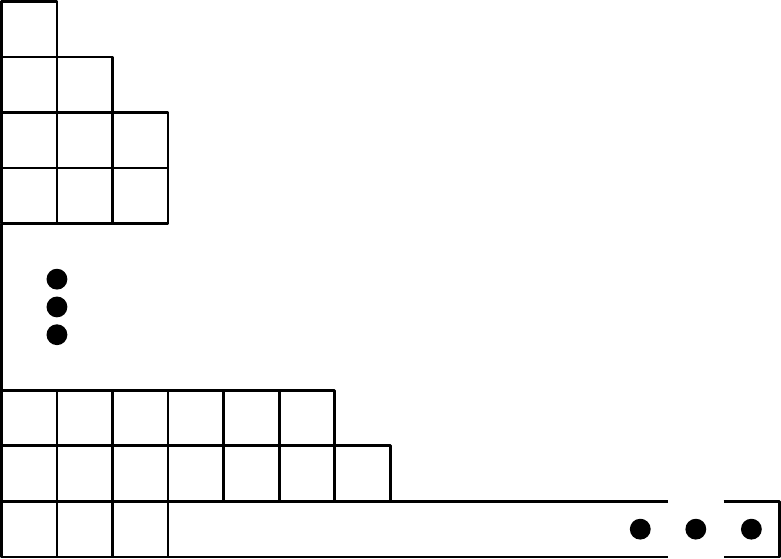}
\caption{A PRP obtained by decoupling punctures with symmetries $U(1)^{n_i} $. It has $L_i-L_{i+1} \le 1$ for $i>1$.}
\label{fig:PRP_from_U1s_only}
\end{figure}

Having or not a decoupled field theory on $AdS_5$, many of the 4d theories do not have a dual string or M-theory on a non-singular background which is weakly curved. 
The symmetry of a four dimensional theory having such a dual should be a product of $U(1)$'s. In such a case, all the punctures in $P(G)$ in \autoref{fig:4d_part_correspondence} must have $P_i - P_{i+1}=0,1$. Using the diagrammatic method, the resulting puncture with symmetry $G$ necessarily has $L_i-L_{i+1}=0,1$ for $i>1$ but not for $i=1$, see \autoref{fig:PRP_from_U1s_only}. Conversely, not every puncture of the form shown in \autoref{fig:PRP_from_U1s_only} can be gauged such that at the other side of the tube there is a sphere with punctures having symmetries $U(1)^{n_i} $ only. This is easily seen by considering \autoref{fig:diagrammatic_decoupling}, since it is not always possible to partition $L_1$ into $P^1_1, P^1_2, \dots, P^1_{\alpha +1} \ge L_2$ with $P_i-P_{i+1} \le 1$. Only surfaces with punctures for which this is possible, might have boundary conditions such that the resulting theory has a gravitational dual which is weakly curved and non-singular. These boundary conditions are a generalization of \autoref{fig:4d_part_correspondence} in which each such puncture is connected through a tube to a sphere with punctures having a symmetry $U(1)^{n_i} $.

Concentrating on a single puncture as in \autoref{fig:4d_part_correspondence}, note that the case from the beginning of this section in which we have only simple punctures on $P(G)$ and $L$ is an L-shaped puncture is special in the following sense.
Assuming that $L$ is not a full puncture, it is the only case in which the rank of the gauge symmetry $G$ in the bulk is equal to the rank of the introduced abelian global symmetry group arising after gauging $G$ in a particular choice of boundary conditions.
This means in the context of \autoref{fig:4d_part_correspondence}, that for $L$ which is not a full puncture being replaced by the punctures $P^i$ of symmetry $U(1)^{n_i} $ each, the case of simple punctures giving an L-shaped puncture is the only one in which $\rk G(L)= \sum _i \rk G(P^i)$. This is a simple consequence of \autoref{subsection:bound_sym_PRP}. Since $L$ is not a full puncture, $\rk G(L)\le N-2$ and we can choose the first argument of the minimum in both \eqref{eq:bound_rk_l_1} and \eqref{eq:bound_rk_l_2}. It then follows that all $P^i_2\le 1$, and to have symmetries of $U(1)^{n_i} $ all the $P^i$ must be simple punctures (and from \eqref{eq:general_PRP_equation}, the resulting $l_k$ is $(1,2, \dots ,m,m, \dots)$, an L-shaped puncture).

\section*{Acknowledgments}

We are thankful to Ofer Aharony for suggesting this project and for useful discussions.
This work was supported in part by the I-CORE program of the Planning and Budgeting
Committee and the Israel Science Foundation (grant number 1937/12), by an Israel Science
Foundation center for excellence grant, by the Minerva foundation with funding from the
Federal German Ministry for Education and Research, by a Henri Gutwirth award from
the Henri Gutwirth Fund for the Promotion of Research, and by the ISF within the ISF-UGC
joint research program framework (grant no. 1200/14).

\appendix
\section{Decoupling on general surfaces through the curve} \label{sec:appendix_g_ge1_analysis}

In this appendix we generalize the analysis of the curve in a decoupling that was done for a sphere in \autoref{subsection:decoupling_sphere} to any surface. The result is used in \autoref{subsection:decoupling_g_ge1}. 

The $\phi _k$ appearing in the SW curve are no longer of the form \eqref{eq:sphere_curve}. Despite this, we are now interested in the region of small $z$, which captures the punctures on the right and the tube (see \autoref{fig:g_ge1_decoupling}). The local behavior of $\phi _k$ in this region is independent of what happens far away, as well as of the genus of the surface. The poles and the zeroes in this region fix it. If we bring $n'_k$ of the zeroes of $\phi _k$ to 0 while taking $w \to 0$, then we have
\begin{equation} \label{eq:general_genus_decoupling_region_SW_curve}
\phi _k \approx u_k \frac{\prod _1^{n'_k} (z-z_i^{(k)}) }{\prod (z-\alpha _i)^{p^i_k} } dz^k
\end{equation}
with $n'_k \le n_k$ and
\begin{equation}
n_k=\Delta _k^L+\Delta _k^R + g(2k-1) .
\end{equation}
For $\Delta _k^R\ge 0$ we ask, as for the sphere, whether we can "flatten" $\Delta _k^R$. On a $g \ge 1$ it is always possible to do that : $n_k -\Delta _k^R \ge \sum _i q_k^i + k-1 \ge 1$.  We argued that on the sphere, if $n_k-\Delta _k^R<0$ (and then could not flatten $\Delta _k^R$, and had to take $u_k \to 0$ as $w \to 0$) or if $n_k-\Delta _k^R=0$, we got $\phi _k^L=0$ which fixed $l_k$ through \eqref{eq:graded_dimension}. It will be useful to reproduce this case in the language we will use. \\
The case $\Delta _k^R \ge 0$ and $n_k-\Delta _k^R \le 0$ is then possible only for the sphere. We have $\Delta _k^R \ge 0$ and $\Delta _k^L \le 0$, and the curve for this case is \eqref{eq:sphere_curve}. In the region of small $z$, it is approximated by
\begin{equation} \label{eq:sphere_decoupling_DR_positive_small_z}
\phi _k \approx \frac{a_0 + a_1z + \dots + a_{n_k} z^{n_k} }{\prod (z-\alpha _i)^{p_k^i} } dz^k  .
\end{equation}
For $w \ll |z| \ll 1$,
\begin{equation}
\phi _k \approx \frac{a_0 +a_1z + \dots +a_{n_k} z^{n_k} }{z^{\Delta _k^R+k} } dz^k .
\end{equation}
In the limit $w \to 0$, this region becomes the region of $l_k$. For this we need to decouple the tube completely. For $\Delta _k^L=0$, $a_{n_k=\Delta _k^R} $ becomes a Coulomb branch parameter for the gauge group along the tube. To decouple the tube, we take $a_{\Delta _k^R} =0$ (not just approach 0 as $w \to 0$, but equal to zero). For $\Delta _k^L<0$ we do not have a parameter for the gauge group for this $k$. In both cases we are left with possible non-zero $a_i$ for $i \le \min(n_k,\Delta _k^R-1)=n'_k$. We must take all the remaining $a_i \to 0$ as $w \to 0$, because otherwise we would get $l_k >k$. This fixes $\phi _k^L=0$ for this $k$. \\
To recover $r_k$ we do the following. To describe the decoupling RHS sphere we change variables $z=yw$, so
\begin{equation}
\phi _k \approx w^{-\Delta _k^R} \cdot \frac{a_0 + a_1 yw+ \dots +a_{n'_k} y^{n'_k} w^{n'_k} }{\prod (y-\alpha '_i)^{p_k^i} } dy^k
\end{equation}
where $\alpha _i=w\alpha '_i$. $\phi _k$ in the $y$ coordinate should be finite to describe the RHS sphere (we have an equation for $\lambda =xdz$ which should give finite cycles; the integration in $y$ should be finite). Therefore $a_i= w^{ \Delta _k^R-i} a'_i$ as $w \to 0$ (they indeed go to 0 as we demanded since $\Delta _k^R - n'_k > 0$). The resulting curve of the decoupling sphere is
\begin{equation} \label{eq:sphere_decoupling_RHS_curve_DR_positive}
\phi _k = \frac{a'_0 + a'_1 y + \dots +a'_{n'_k} y^{n'_k} }{\prod (y-\alpha '_i)^{p^i_k} } dy^k .
\end{equation}
The pole order at infinity is obtained as usual by a change of variable $y=1/y'$ with $y' \to 0$ as $y \to \infty $. This gives the pole structure of $r_k$ : $r_k=k-\Delta _k^R + n'_k$. For $\Delta _k^L<0$ this is $r_k=k-\Delta _k^R+n_k = k+\Delta _k^L=\sum _i q_k^i$. This is what we are familiar with, when looking from the LHS point of view having $\Delta _k^L<0$ where we said we do not need to do anything, giving $r_k=\sum _i q_k^i$. For $\Delta _k^L=0$, $r_k= k-1$, again as we are familiar with. Note, as stated before, if $\Delta _k^R=0$, then in our case either $\phi _k \approx a_0 / z^k \cdot dz^k$ or $\phi _k=0$ to begin with. In both cases, turning off the tube, $\phi _k^R=0$ fixing $r_k$ from $d_k^R$, $r_k=k-1$.

Now return to surfaces of general genus. Suppose that $\Delta _k^R \ge 0$ and $n_k-\Delta _k^R \ge 1$. As mentioned, the first condition implies the second in $g \ge 1$ surfaces. Meanwhile assume $\Delta _k^R>0$. In the region of small $z$, where the right punctures approach each other, we have the curve behavior for any surface \eqref{eq:general_genus_decoupling_region_SW_curve} : 
\begin{equation} \label{eq:approx_phi_k_decoupling_sphere}
\begin{split}
\phi _k & \approx \frac{a_0+a_1z+ \dots + a_{n'_k}  z^{n'_k} }{\prod (z-\alpha _i)^{p_k^i} } dz^k = \\
&= w^{-\Delta _k^R} \cdot \frac{a_0+a_1 wy + \dots +a_{n'_k}  (wy)^{n'_k} }{\prod (y-\alpha '_i)^{p_k^i} } dy^k .
\end{split}
\end{equation}
Again $\phi _k$ in the $y$ coordinate should be finite since it describes the RHS sphere. It means we should take $a_i = w^{\Delta _k^R-i} a'_i$ as $w \to 0$. 
\\
The leading behavior of $\phi _k$ in the region $w\ll |z| \ll$ other poles and zeroes, is (using $z \gg z_i^{(k)} ,\alpha_i$)
\begin{equation}
\phi _k \approx \frac{ a_{n'_k}  z^{n'_k} + \dots}{z^{\Delta _k^R+k} } dz^k .
\end{equation}
The higher order in $z$ terms in "..." are finite in the decoupling limit in general (they can go to zero, but this corresponds to further taking zeroes of $\phi _k$ in the rest of the surface to 0, but still much larger than $w$). There will be at least one such term for the choice of $n'_k$ we will do. This form fixed for us $l_k$. Since we cannot have $l_k>k-1$ we take $n'_k =\Delta _k^R-1$. $a_{n'_k} \to 0 $ necessarily as we saw, so there is no $1/z^{k+1} $ term in $l_k$. The Coulomb branch parameter of the gauge group along the tube is taken to zero. We therefore get $l_k=k-1$ (the value of $n'_k$ we have taken, satisfying $n_k-n'_k>0$, ensures we have higher orders in $z$, so $\phi _k^L \neq 0$). For $r_k$ we use the curve on the right sphere
\begin{equation}
\phi _k = \frac{a'_0+a'_1 y + \dots + a'_{\Delta _k^R-1} y^{\Delta _k^R-1} }{\prod (y-\alpha '_i)^{p^i_k} } dy^k
\end{equation}
(an index of $k$ is omitted from $a_i$ and $a'_i$ for clarity of the equations, but it should be kept in mind that these $a$'s are different for different $k$'s). 
Note $ \frac{a_i}{a_{n'_k} } \sim w^{n'_k-i} $ as indeed expected since the zeroes in $ a_{n'_k} \prod (z-z_i^{(k)} )$ are proportional to $w$. 
$r_k$ can be read from the pole at infinity and is $r_k=k-\Delta _k^R+\Delta _k^R-1 = k-1$.

Note we have taken $n'_k=\Delta _k^R-1$. We could not take smaller $n'_k$ because it would give $l_k>k-1$ as mentioned, but we could ask why not take $n'_k >\Delta _k^R$ (note $n'_k=\Delta _k^R$ is the same as what we did before since we take its coefficients to be 0, not affecting either curve). The reason is that in the last equation we would get in the numerator terms up to $y^{n'_k} $ giving $r_k=k-\Delta _k^R+n'_k>k$. This justifies "not flattening further".

Regarding $\Delta _k^R=0$, we must take $n'_k=0$ for the same reason. We still have $l_k=k-1$. In the SW curve of the RHS sphere we only have $a'_0$ in the numerator. We should take it to 0 in order to decouple the tube, giving $\phi _k^R=0$ as we already know. This fixes $r_k=d_k^R-\Delta _k^R+k-1=k-1$.

Lastly, in $g \ge 1$ surfaces we need also to consider $k$'s for which $\Delta _k^R<0$ and $n_k-\Delta _k^R \ge 1$. We said we do not need to bring any $z_i^{(k)} $'s to 0, that is $n'_k=0$. The curve in the region $w\ll |z| \ll 1$ is then
\begin{equation} \label{eq:general_g_DR_negative_tube}
\phi _k \approx a_0 \frac{1}{z^{\sum_i p_k^i} } dz^k .
\end{equation}
There are no Coulomb branch parameters for the gauge group along the tube. This gives $l_k= \sum_i p_k^i$. In the region of the decoupling sphere we have
\begin{equation}
\phi _k \approx \frac{a_0}{\prod (z-\alpha _i)^{p^i_k} } dz^k = w^{k-\sum _i p^i_k} \frac{a_0}{\prod (y-\alpha '_i)^{p^i_k} } dy^k \xrightarrow{w \to 0} 0
\end{equation}
since $\sum _i p_k^i < k$ and $a_0$ cannot diverge because of \eqref{eq:general_g_DR_negative_tube}. For the RHS curve we get $\phi _k^R=0$ fixing $r_k=d_k^R-\sum _i p_k^i +2k-1=2k-1-\sum_i p_k^i$. \\
We could as before wonder what happens if we were taking $n'_k>0$ $z_i^{(k)} $'s to 0, giving $l_k=\sum_i p_k^i - n'_k < \sum _i p^i_k$. Again, this is the question of why not "flatten further" in our previous considerations. The RHS curve would then be \eqref{eq:approx_phi_k_decoupling_sphere}. The parameters in the resulting RHS curve should go to 0, otherwise we would get $r_k>k$. Then $\phi _k^R=0$ and still $r_k=2k-1-\sum_i p_k^i$, but now $l_k< \sum _i p_k^i$. It does not give a new decoupling from the following reason. We took $n'_k$ parameters to 0 as $w \to 0$ but we do not see them in the RHS SW curve. It seems as if the number of parameters is not conserved. To understand what happens, return to the first situation we analyzed --- the sphere with $\Delta _k^R>0$ and $\Delta _k^L<0$. This is just switching left and right of this situation, on the sphere. The curve was approximated by \eqref{eq:sphere_decoupling_DR_positive_small_z} for small $z$ there, corresponding to non-small $z$ here. Write $a_0+a_1z+\dots +a_{n_k} z^{n_k} =a_{n_k} \prod _1^{n_k} (z-z_i)$. If all the zeroes $z_i$ behave as $z_i \sim w^1$ and $a_{n_k} \sim w^{\Delta _k^R-n_k} $, we will get $a_i \sim a_{n_k} z_i^{n_k-i} \sim w^{\Delta _k^R-i} $ exactly as we obtained there. Now suppose we want to lower the degree of the polynomial in the numerator of \eqref{eq:sphere_decoupling_RHS_curve_DR_positive}, which is exactly what we are doing here (less parameters in the $\Delta >0$ side). Let us lower it by one for instance. For this we need $a'_{n_k} \to 0$, or $a_{n_k} \sim w^{\Delta _k^R-n_k + 1} $ (or $w^{\Delta _k^R-n_k+s} $ with $s>0$). To preserve $a'_{n_k-1} $ finite, we need $a_{n_k-1} $ to still behave as $w^{\Delta _k^R-n_k+1} $. This requires that at least one $z_i$ will behave as $w^0$ (or $w^{1-s} \gg w$). We get a $z_i$ that stays in the LHS of the analysis there, which corresponds to the RHS in our current analysis (having $\phi _k=0$). We conclude that getting $l_k< \sum _i p_k^i$ in our case, is just like taking $a'_i$'s to 0 in \eqref{eq:sphere_decoupling_RHS_curve_DR_positive}, which is a degenerate case of the general $l_k = \sum _i p_k^i$ situation; we just take some Coulomb branch parameters to 0 there.

To conclude, for $g \ge 1$ surfaces, we obtained that for $\Delta _k^R \ge 0$, $l_k=r_k=k-1$, while for $\Delta _k^R<0$, $l_k=\sum _i p_k^i$ and $r_k=2k-1-\sum _i p^i_k$. There was a single tube Coulomb branch parameter of dimension $k$ for $\Delta _k^R \ge 0$. The punctures that decouple alone fix the gauge group along the tube and the two punctures that are created. The two possibilities for $\Delta _k^R$ of the right side of \autoref{fig:g_ge1_decoupling} are just as the ones on the right of figures \ref{fig:two_spheres_SUSU},\ref{fig:two_spheres_SUSp1}. The resulting gauge group, $r_k$ and $l_k$ are just as there.

\bibliographystyle{/Users/vladinaro/Desktop/University/utphys2.bst}
\bibliography{/Users/vladinaro/Desktop/University/Bibliography}

\end{document}